\documentclass[sigconf, nonacm]{acmart}

\usepackage{amsmath}
\usepackage{amsfonts}
\usepackage{amsmath}
\usepackage{amsthm}
\usepackage{graphicx}
\usepackage{textcomp}
\usepackage{xcolor}
\usepackage{enumitem}
\usepackage{subcaption}
\usepackage{amsmath, amsthm,natbib}
\usepackage{graphicx}
\usepackage{amsfonts}
\usepackage{rotating}
\usepackage[normalem]{ulem}
\usepackage{algorithm}
\usepackage[noend]{algpseudocode}
\algrenewcommand{\Return}{\State\algorithmicreturn~}
\usepackage{xspace}
\newcommand{\AC}[1]{{\color{red}[AC: #1]}}

\newcommand{\delete}[1]{{\color{green}\ignore{#1}}}
\newcommand{\revision}[1]{{\color{black} #1}}

\def\LL{\mathcal{L}}

\def\CC{\mathcal{C}}

\def\CT{\ensuremath{T}}

\def\probname{MVR-P}

\newcommand{\ccost}{\delta}
\newcommand{\cost}[1]{\delta_{#1}}
\newcommand{\storage}[1]{sz_{#1}}
\newcommand{\ignore}[1]{}

\newcommand{\opcost}{\ensuremath{\text{cost}^*}}
\newcommand{\bcalgo}{Parent Choice}
\newcommand{\bcalgopc}{ParentChoice}
\newcommand{\bccost}{\ensuremath{\text{cost}}}
\newcommand{\totcost}{\ensuremath{\Delta}}
\newcommand{\decprob}{\ensuremath{RP}}
\newcommand{\bSin}{\ensuremath{S}}
\newcommand{\bSu}{\ensuremath{S_{+u}}}
\newcommand{\children}{\ensuremath{Children}}
\DeclareMathOperator*{\argmax}{arg\,max}

\newcommand{\bba}{\ensuremath{\mathbf{a}}}

\newcommand{\bbc}{\ensuremath{\mathbf{c}}}

\newcommand{\bbf}{\ensuremath{\mathbf{f}}}

\newcommand{\bbi}{\ensuremath{\mathbf{i}}}

\newtheorem{definition}{Definition}
\newtheorem{problem}{Problem}
\newtheorem{theorem}{Theorem}
\newtheorem{proposition}{Proposition}

\newcommand{\TOOLNAME}{{\textbf{CHEX}}\xspace}
\newcommand{\toolname}{\texttt{\textbf{CHEX}}\xspace}

\def\BibTeX{{\rm B\kern-.05em{\sc i\kern-.025em b}\kern-.08em
    T\kern-.1667em\lower.7ex\hbox{E}\kern-.125emX}}

\newcommand{\mysubfigthreebox}[9]
{
\def\tempa{#1}
\def\tempb{#2}
\def\tempc{#3}
\def\tempd{#4}
\def\tempe{#5}
\def\tempf{#6}
\def\tempg{#7}
\def\temph{#8}
\def\tempi{#9}
\mysubfigthreeboxcont
}

\newcommand{\mysubfigthreeboxcont}[2]
{
\begin{figure*}
        \centering
        \begin{tabular}{ccc}
        \begin{subfigure}[b]{0.28\textwidth}
                \centering
                {%
                {\includegraphics[width=0.8\textwidth]{\tempa}}%
                }%
                \vspace{-5pt}
                \caption{\tempb}
                \label{\tempc}
        \end{subfigure}%
        &
        \begin{subfigure}[b]{0.28\textwidth}
                \centering
                {%
                {\includegraphics[width=0.8\textwidth]{\tempd}}%
                }%
                \vspace{-5pt}
                \caption{\tempe}
                \label{\tempf}
        \end{subfigure}
        &
        \begin{subfigure}[b]{0.28\textwidth}
                \centering
                {%
                {\includegraphics[width=0.8\textwidth]{\tempg}}%
                }%
                \vspace{-5pt}
                \caption{\temph}
                \label{\tempi}
        \end{subfigure}
        \end{tabular}
        \vspace{-10pt}
        \caption{#1}\label{#2}
        \vspace{-10pt}
\end{figure*}
}

\algrenewcomment[1]{\(\triangleright\) {\scriptsize #1}}
\algnewcommand{\LineComment}[1]{\State \(\triangleright\) {\tiny #1}}

\begin{document}
\title{CHEX: Multiversion Replay with \\ Ordered Checkpoints}

\author{Naga Nithin Manne}
\authornote{Work done as part of a summer internship at DePaul.}
\affiliation{%
  \institution{Argonne National Lab.}
  \streetaddress{9700 S. Cass. Ave}
  \city{Lemont}
  \state{IL}
  \country{USA}
}
\email{nithinmanne@gmail.com}

\author{Shilvi Satpati, Tanu Malik}
\affiliation{%
  \institution{DePaul University}
  \streetaddress{243 S. Wabash Ave.}
  \city{Chicago}
  \state{IL}
  \country{USA}
}
\email{ssatpati,tanu.malik@depaul.edu}

\author{Amitabha Bagchi}
\affiliation{%
  \institution{IIT, Delhi}
  \streetaddress{1 Th{\o}rv{\"a}ld Circle}
  \city{Delhi}
  \country{India}
}
\email{bagchi@cse.iitd.ac.in}

\author{Ashish Gehani}
\affiliation{%
  \institution{SRI}
  \city{Menlo Park}
  \state{CA}
  \country{USA}
}
\email{ashish.gehani@sri.com}

\author{Amitabh Chaudhary}
\orcid{0000-0001-5109-3700}
\affiliation{%
  \institution{The University of Chicago}
  \city{Chicago}
  \state{IL}
  \country{USA}
}
\email{amitabh@uchicago.edu}





\begin{abstract}
    In scientific computing and data science disciplines, it is often necessary to share application workflows and repeat results. Current tools \delete{enable researchers to encapsulate and }containerize \delete{their }application workflows\revision{, and share the resulting container }\delete{into a package that can be shared }for repeating results. These tools\revision{, due to containerization, do} improve sharing of results\delete{,}\revision{. H}owever, they do not improve the efficiency of replay. In this paper\revision{,} we present the multiversion replay problem which arises when \delete{ the researcher containerizes multiple versions of an application, and a collaborator wishes to replay the versions efficiently.}\revision{ multiple versions of an application are containerized, and each version must be replayed to repeat results.}
 To avoid executing each version separately, we develop \texttt{\TOOLNAME}\revision{, which checkpoints program state and determines when it is permissible to reuse program state across versions. It does so using system call-based execution lineage.}\delete{ which enables sharing of computations across versions. \texttt{\TOOLNAME} uses system call-based execution lineage to determine points where execution of any two versions is equivalent and from where it branches creating an execution tree across multiple versions.} Our capability to identify common computations across versions enables us to consider optimizing replay using an in-memory cache, based on a checkpoint-restore-switch system.
    We show the multiversion replay problem \revision{is} NP-hard, and propose efficient heuristics for \revision{it}. 
    \texttt{\TOOLNAME} reduces overall replay time by sharing common computations but avoids storing a large number of checkpoints. We demonstrate that \texttt{\TOOLNAME} maintains lightweight package sharing, and improves the total time of multiversion replay by 50\% on average.
\end{abstract}

\ignore{


\begin{abstract}
Common computations across multiple versions of a program are often shared to avoid recomputation costs. Current approaches for sharing rely on maintaining computational state at either the function or the program level. Effective state reuse, however, requires accurate program and data dependency analyses which is often cumbersome to obtain and becomes a data management challenge. In this paper, we develop \texttt{\TOOLNAME}, a system for multiversion replay, which replays multiple versions of a program, and uses checkpointing as a method for sharing common computations across versions. 
\texttt{\TOOLNAME} reduces overall replay time by sharing common computations but avoids storing a large number of checkpoints. \texttt{\TOOLNAME} relies on REPL-style programs and audited lineage during version execution to determine common checkpoint locations. However, as we show, deciding checkpoint locations for efficient replay in space-constrained setting is NP-hard. Therefore \texttt{\TOOLNAME} uses two efficient depth-first-based heuristics for deciding checkpoint locations, one that is time-efficient and another that is space-efficient. We experiment with \texttt{\TOOLNAME} on real-world machine learning and scientific computing notebooks and synthetic datasets, and show that \texttt{\TOOLNAME} improves total replay time of real and synthetic applications by 50\%. 
\end{abstract}
}

\maketitle

\section{Introduction}
\label{sec:introduction}

Suppose that Alice is researching different image classification pipelines. She has a large labeled set of images and progressively tries different combinations of preprocessing steps and neural network architectures. \delete{Occasionally, she fine tunes a hyperparameter.} For example, she may replace an entire step with one that is more sophisticated but slower, or vice-versa. \revision{As she makes changes, she keeps a copy of the previous versions in separate Jupyter notebooks.  We call these her different \emph{program versions;} they are similar to different \emph{experiments in scientific computing.}} Once done, Alice would like to share her different program versions
with Bob so that he can \delete{verify them and reproduce}\revision{independently repeat and regenerate} the results and verify Alice's work. We say Bob faces the \textit{multiversion replay problem}: executing all the versions given to him by Alice as efficiently as possible where (i) many versions repeat some of the same preprocessing steps, but (ii) without reusing any of Alice's own computation. In this paper, we address this problem.

Collaborative scenarios such as the one \delete{mentioned}above arise routinely in scientific computing and data science, where sharing, repeating and verifying results is \delete{necessary}\revision{common}. \delete{Some such scenarios are: model selection and management~\cite{miao2017modelhub,vartak2018mistique}, reproducible analysis~\cite{peng2011reproducible,stodden2016enhancing,freire2018reproducibility}, causality inference~\cite{kwon2016ldx,Kwon18} and software repair~\cite{monperrus2018automatic}. Consequently, s}\revision{S}everal tools have been recently proposed for it~\cite{Janin:CARE,Chirigati:2016:ReproZip,o2017engineering,Pham:2013:PTU,Pham:ICDE:LDV,Pham:VLDB:LDV,SciunitI:2017:sciunit-run,Yuan:2018:Sciunit}.\delete{, including a couple by one of the authors} These tools audit \revision{the execution of a program,}\delete{an application} and create a \revision{container-like}\delete{self-contained} package consisting of \revision{all}\delete{application} files \revision{referenced by the program during its execution.} \delete{and execution lineage.}\revision{This package can then be used to repeat results in different environments.}
\revision{These tools have much to offer; they do not, however, exploit the efficiency possible by solving the multiversion replay problem. As the reproduction of results becomes increasingly time consuming, addressing such problems is critical.}
\delete{Despite the fact that executing multiple versions of a program can be very time-consuming, especially when programs are compute and data-intensive, we find that environments enabling {\em efficient} multiversion \revision{replay}\delete{(\textit{re-})execution} are currently not reported in the literature.}

\revision{One of the above tools is the Sciunit system, developed by some of the co-authors.}\delete{The motivation for this problem emerged during the development of the Sciunit system~\cite{Ton:2017:Sciunit,SciunitI:2017:sciunit-run} by one of the authors of this paper. Sciunit enables} \revision{It allows} multiple versions of a program to be included in the same package and shared for repetition~\cite{Ton:2017:Sciunit,Nakamura:2020:HiPC}. We noted that having two or more versions in the same package sets up a natural opportunity for reusing computations that are often common across versions---\textit{i.e.,} a number of versions may perform the same computations for quite some time before the\revision{y}\delete{development process} branch \revision{out}\delete{es} as the \revision{researcher}\delete{programmer} tries out different options.  \revision{But}, \revision{i}\delete{I}n order to accurately identify computations that can be reused across versions, we need to be able to determine the point to which the execution of two versions can be treated as equivalent and from \revision{which they}\delete{where it} branch\delete{es}. We \revision{develop}\delete{build on execution lineage inference methods developed earlier~\cite{Pham:ICDE:LDV,Nakamura:2020:TaPP} and create} a methodology to identify common computations \revision{in program code fragments or \textit{cells} of the }versions. \revision{This methodology depends on lineage audited during program execution~\cite{Pham:2013:PTU,Pham:ICDE:LDV,Nakamura:2020:TaPP}.} \delete{Indeed, it is our capability to identify common computations across versions that allows us to focus on efficient repetition in the paper.}
\revision{For repetition, at Bob's end, we}\delete{In particular, our approach is to} share computational state across versions in the form of checkpoints\revision{.}\delete{during the repetition process.} \revision{Let us now see with an example how sharing computational state across versions via checkpoints creates an opportunity to optimize computational time when repeating multiple versions.} \delete{that can be solved efficiently by using  caching. Once this is made possible, we are able to state an optimization problem that can be solved to use caching to efficiently repeat the multiversion program.

Let us see with an example how sharing computational state across versions enables efficient multiversion replay.}

\begin{figure}[h]
    \centering
    \includegraphics[viewport=0mm 0mm 200mm 85mm,width=0.8\linewidth]{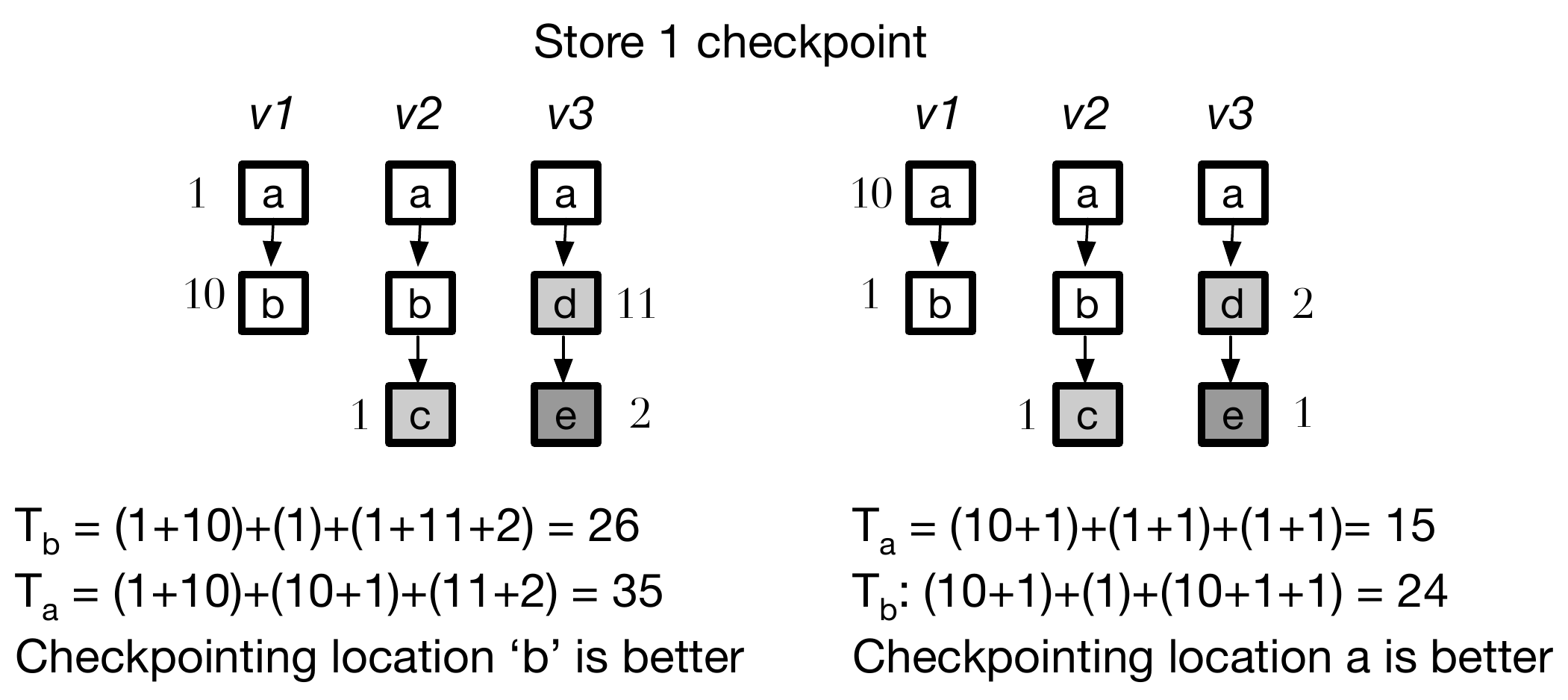}
    \caption{\revision{Deciding  checkpoint location depends on cost and size estimates of the cells. }}
    \label{fig:optimization}
\end{figure}

\delete{Suppose that Bob wants to replay Alice's shared package consisting of three versions.} \revision{Suppose that Alice has shared with Bob a package with three versions.} \delete{Let us a} Assume Alice \delete{is developing} \revision{has developed her code} in the REPL style and her first version is divided into \revision{two}\delete{three} cells that take time 1 minute \revision{ and }\delete{,} 10 minutes\delete{, and 1 minute}, respectively (Figure~\ref{fig:optimization} left). Her second version has the same first two cells but \revision{she adds}\delete{changes} \revision{a}\delete{the} third cell. Her \revision{last}\delete{third} version, which \revision{processes}\delete{changed} the dataset, has the same first cell, but \revision{diverges}\delete{changes} after that, and \revision{cells 2 and 3} now takes 11 and 2 minutes, respectively. Now, during repetition, if Bob  checkpoints cell $b$ while computing $v1$ and then restores that checkpoint for $v2$, he can complete $v2$ in just 1 unit of time (saving 11 units in $v2$). This checkpoint is\revision{, however, not useful}\delete{useless} for $v3$ since cell $b$ has been changed to \revision{$d$}\delete{$e$} in this version. \revision{Observe,}\delete{It is easy to see} that although $a$ is common across all three versions, checkpointing $a$\revision{, instead of $b$, }is not optimal. In the example on the right, on the other hand, checkpointing $a$ is the better option since the bulk of the computation takes place in $a$.
The example above leads to the question: {\em Why doesn't Bob just checkpoint both $a$ and $b$?} Storage is cheap after all! Indiscriminate caching, however, is not a practical solution: In machine learning examples, e.g.,  checkpointing all cells across versions (as our  experiments indicate) can lead to a memory requirement in the range of 50-550GB, for even moderately-sized multiversion programs. Thus, we consider a limited in-memory space as this avoids additional I/O costs from checkpointing. \revision{ Limited space }\delete{this} leads to a\revision{n optimization problem}\delete{computational bottleneck}: As we will show, given limited space and multiple versions with different cost and size estimates of cells, deciding checkpoint locations for efficient replay of all versions is an NP-hard problem. We present efficient heuristics to solve this problem.

\paragraph{The \texttt{\TOOLNAME} system}
\delete{The big picture is this:} We present \texttt{\TOOLNAME}, a system for efficient multiversion replay that uses recorded lineage shared in \revision{container-like auditing systems}\delete{self-contained packages} to (i) determine when \revision{the program}\delete{computational} state is identical across versions, and (ii)
decides which common computations to save in an in-memory, limited-size cache, and where to continue recomputing. Effectively, \texttt{\TOOLNAME} computes an efficient plan for Bob to use his cache to repeat Alice's multiversion program with minimum computation cost. Subsequently \texttt{\TOOLNAME} repeats the computation according to this plan.

To execute the multiversion program, we first need a plan for sharing and reusing computational state across versions. A possible approach would be to reuse program elements such as the output of functions, expressions, or jobs. \revision{Such reuse approaches were examined in ~\cite{Guo:TaPP:2010:IncPy,Guo:ISSTA:2011:IncPy, Nectar}. However,}\delete{
For instance, IncPy~\cite{Guo:TaPP:2010:IncPy,Guo:ISSTA:2011:IncPy} avoids recomputing a function with the same inputs when it is called repeatedly or across program versions. Similarly, in Nectar~\cite{Nectar}, intermediate datasets are cached, and instead of running the same program with given inputs, the datasets are reused. However, neither of these approaches work for us. To share state,} these methods make  assumptions about the programs---\revision{they}\delete{both methods} are limited to programs with no side-effects and apply to specific (functional or interpreted) programming languages. Such assumptions are too restrictive in a sharing scenario.  


Our approach is program-agnostic and we, instead,  use checkpoints. A checkpoint saves the computational state 
at a specific program location so that \revision{the \textit{same} program} can be restored from the location at a later time. To share computations, we extend \textit{checkpoint-restore} to \textit{checkpoint-restore-switch}, in which a system checkpoints a common computational state and and restores it later to resume a {\em different} version of the program.
\ignore{Figure\revision{~\ref{fig:notebook}} shows \revision{two programs which have common code from start till they load the training data set \texttt{new\_fashion}. If the first program is checkpointed after this instruction, the  checkpoint can reused executing the remaining of the second  program. }}

The challenge \revision{of checkpoint-restore-switch, however, is determining locations at which to checkpoint, since ideally programs may be checkpointed after each instruction.}
\delete{sharing computational state across two program versions; it restores a checkpoint of the shared computational state from the first version and switches to resume execution of the second version at the marked location. 
The advantage of checkpointing is that if the parameter to function \texttt{stageModelTrain} is not 
changed, the computation up to that location is reusable for \texttt{stageModelTest} in the second version.} \delete{Such reuse is often not achievable with IncPy and Nectar due to inherent randomization in the computation that is common to machine learning-based functions, as in the example here.} 
\delete{Checkpoint-restore-switch brings its own set of challenges. In general, there are several program locations at which to checkpoint.}Even if we decide at a fixed number of program locations, before reusing a checkpoint we must verify that two versions share the same computational state at a given program location. \revision{{\em In this paper we solve this dual challenge by showing that when a program is divided into {\em cells}, computational state can be shared across versions by using fine-grained execution lineage}}. Dividing a program into cells is used in read-evaluate-print (REPL)  programming environments, which are increasingly popular~\cite{Guo2020design}. However, \texttt{\TOOLNAME} does {\em not} necessitate that a user employ REPL style; it  transparently divides a program into REPL-style cells. \delete{Unlike fine-grained lineage support during interactive development~\cite{macke2020fine}, \texttt{\TOOLNAME} uses system call execution lineage to combine versions into a data structure called the \textit{execution tree}. This execution tree represents the development process in a natural way, branching when the developer decides to change a cell. Each cell becomes a vertex of the execution tree and each root-to-leaf path of the execution tree represents one version. The vertices of this tree are annotated with compute time and checkpoint size estimates. \texttt{\TOOLNAME} uses these estimates and the structure of the tree to decide heuristically where to checkpoint and where to recompute.}

This paper contributes the following:

\noindent {\bf Maintains lightweight package sharing.} \texttt{\TOOLNAME} does not require users like Alice to share checkpoints as part of the shared package. Instead, \texttt{\TOOLNAME} audits the execution of each version to \revision{record execution details}\delete{create an enhanced specification}. We note that in reproducibility settings, it is not desirable to allow Alice to share her checkpoints since that defeats the purpose of reproducibility.

\noindent {\bf Merging versions based on lineage.} \texttt{\TOOLNAME} \revision{compares}\delete{uses} fine-grained lineage to check if \delete{at a given cell location} the program state is common across \revision{cell} versions. It combines versions into an \textit{execution tree}. 

\noindent {\bf Deciding checkpoint location.} Given that the multiversion replay problem under space constraints is computationally intractable, we rely on depth-first-search (DFS) traversals of the execution tree to help us identify a subset of possible checkpointing decisions for the execution units of the program. We call the members of this subset {\em DFS-based replay sequences.}
 We propose two heuristic algorithms for deciding which cell state to checkpoint such that the multiple versions can be replayed in a minimum amount of time. 

\noindent {\bf Experiments on real and synthetic datasets.}   We experimented with real machine learning and scientific computing notebooks as well as synthetic datasets, showing that \texttt{\TOOLNAME} improves  the total time of multiversion replay by 50\%, or correspondingly replays twice the number of versions in a given amount of time. \revision{We show that the overheads of creating execution trees is significantly lower than the gain from replay efficiency.}

\noindent {\bf Working prototype system:} We have developed a prototype \texttt{\TOOLNAME} system, which given an execution tree performs multiversion replay. \delete{Our }\texttt{\TOOLNAME} \delete{system} currently uses standard auditing  methods, developed by us, to build execution trees \revision{and determine cell reuse}~\cite{Ton:2017:Sciunit,SciunitI:2017:sciunit-run}. For multiversion replay, we extended these methods to work with interactive Jupyter notebooks, as well as, transform regular programs to REPL-style computation via code-style paragraphs.

\section{Background}
\label{sec:prelim}
\revision{

\begin{figure}[t]
    \centering
    \includegraphics[viewport=0mm 0mm 200mm 135mm,width=\linewidth]{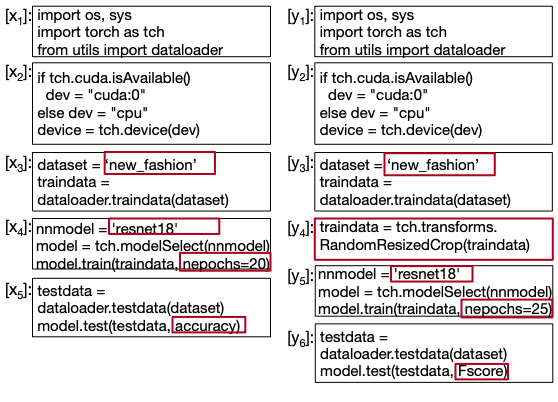}
    \vspace{-10pt}
    \caption{\revision{An illustration of REPL programs.  The left program $L_1$ trains a machine learning model (resnet18) on a training dataset and evaluates its accuracy on a test dataset.  The right program $L_2$ is the same, except that it adds a preprocessing step to the training dataset.  $L_1$ has 5 cells, $L_2$ has 6 cells.}}
    \label{fig:notebook}
\end{figure}

We briefly describe the REPL environment under which \toolname operates. \toolname is not limited to the REPL environment but this is easy to illustrate visually so we adopt this for ease of exposition. We discuss generalization to other environments in Section~\ref{sec:Discussion}. 

REPL or Read-Evaluate-Print-Loop is a programming environment. A popular example is the Jupyter notebook.  As shown in Figure~\ref{fig:notebook}, it contains code partitioned into \emph{cells}.  Developers typically use a separate cell for each ``step" of the program: preprocessing the dataset, training the model, etc.  This allows them to interactively test each step before writing the next.
One restriction is that control flow constructs, such as if-blocks, loops, cannot be split across cells.   We denote a REPL program by an ordered list of cells, \textit{e.g.,} the left program in Figure~\ref{fig:notebook} is denoted $L_1 = [x_1, x_2, \ldots, x_5]$, and the right program as $L_2 = [y_1, y_2, \ldots, y_6]$.  

In a typical REPL execution, cells are executed in sequence from the first to last. While the Jupyter notebook allows out-of-order cell execution, we do not consider such execution. (We elaborate on this constraint further in Section~\ref{sec:Discussion}.) The state of the program at the end or beginning of each cell is termed the program state. The program state at any point of execution consists of the values of all variables and objects used by the program at that point --- intuitively, it is all the contents of the memory associated with the program. So, e.g., for the program $L = [x_1, x_2, \ldots, x_5]$, the corresponding program states are $[ps_0, ps_1, \ldots, ps_5]$, in which $ps_{i-1}$ denotes the program state just before cell $x_{i}$ is executed.  The state $ps_0$, which is just before the first cell is executed, includes the value of the environment and any initial input. 

\toolname works in combination with an {\em auditing system} which monitors executions and provides the following details about each program state, $ps_i$:

\begin{itemize}[leftmargin=*]
\item {\bf computation time}, $\delta_{i}$, the time to reach the program state $ps_{i}$ from its predecessor $ps_{{i-1}}$, 

\item {\bf size}, $sz_{i}$, size of the program state $ps_i$,

\item {\bf code hash}, $h_i$, computed by
hashing code in cell $i$, and


\item {\bf lineage}, $g_i$, which is determined by combining the predecessor cell's lineage with the sequence of system events that are triggered by program instructions in the cell $i$ and the hashes of the associated external data dependencies. 
Thus, $g_i = (g_{i-1}, h_i, E_i)$, where $E_i$ is the ordered set of system events in  cell along with the hash of the content accessed by the event. Initially, $g_0 = \{\}.$
\end{itemize}

To see why $g_i$ is so defined, we note that the execution of the program code in cell $i$ (and the code in previous cells) resulted in $ps_i$. Therefore, $ps_i$ 
at the end of a cell's execution depends on its (i) initial environment, (ii) code that is run, and (iii) external input data. The environment is determined by the execution state at the start of the cell. Thus, (i) and (ii) are captured via $g_{i-1}$ and $h_i$. 
Further, every  external input data file $f$ is accessed via a system call event. For each such event, we record a hash of its contents of $f$ in $E_i$.

Figure~\ref{fig:lineage} shows the audited information for the two programs, $L_1$ and $L_2$. The ordered set of system events for the third cells of the two programs are shown in the shaded box below. 

\begin{figure}[h]
    \centering
    \includegraphics[viewport=0mm 0mm 200mm 155mm,width=0.8\linewidth]{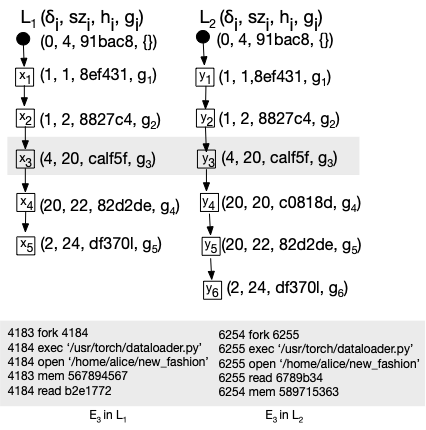}
    \caption{\revision{Auditing of programs $L_1$ and $L_2$ in terms of $\delta$, $sz$, $h$, $g$. The audited lineage of third cell is shown in terms of $E$. We show how to equate this lineage in Section~\ref{sec:chex}.}}
    \label{fig:lineage}
\end{figure}
}

\section{CHEX Overview}
\label{sec:chex-overview}

\begin{figure}[h]
    \centering
    \includegraphics[viewport=0mm 0mm 220mm 55mm,width=\linewidth]{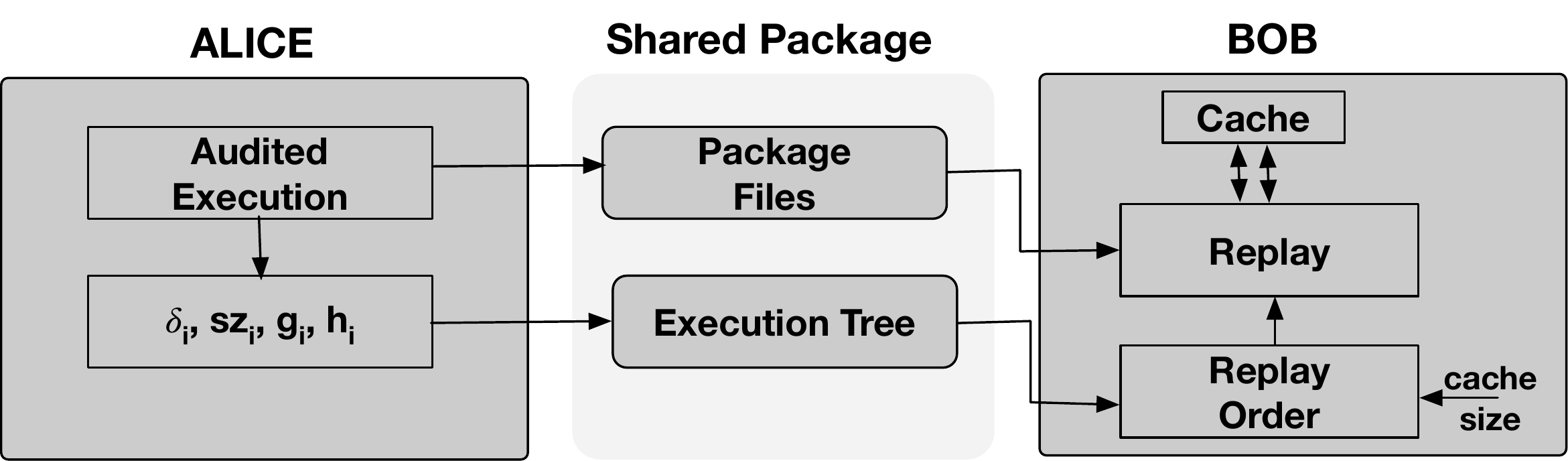}
    \caption{\revision{\texttt{\TOOLNAME} Overview.}}
    \label{fig:chex}
\end{figure}

\revision{As we see in Figure~\ref{fig:notebook}, the two programs behave the same till the end of the third cell ($x_3$ in $L_1$, $y_3$ in $L_2$) and then diverge. If the audited lineage, as shown in Figure~\ref{fig:lineage}, is established to be same, then the program state at the end of $x_3$ can be used before $y_4$. i.e, we can skip executing cells $y_1$ to $y_3$. \toolname uses recorded lineage  to determine when the program state is identical across versions, and decides which common computations to save. We now present a high-level block diagram of  \texttt{\TOOLNAME} in Figure~\ref{fig:chex}.}

 \revision{\texttt{\TOOLNAME has two modes audit and replay. It} is used in audit mode to audit details of executions on Alice's side.}\delete{To begin with Alice uses \texttt{\TOOLNAME} in audit mode  
to execute each version. \texttt{\TOOLNAME} audits details of her executions.} \revision{Details of multiple executions, i.e. the $\delta$, $sz$, $h$ and $g$ of each cell across versions are represented}\delete{It then represents the details of her executions} in the form of a data structure called the \textit{Execution Tree}. \delete{The execution tree contains the computation time and size for each cell.}\delete{The execution tree contains information about the computation cost and average checkpoint size of each cell in each version of Alice's multiversion program. It also contains information about which cells can be identified with each other across versions. This information is encapsulated as a tree structure.} We discuss the execution tree and how it is created in detail in Section~\ref{sec:chex}. \texttt{\TOOLNAME} creates a package of all Alice's versions and their data, binary, and code dependencies, along with the execution tree. This package can now be shared with Bob.

\revision{\texttt{\TOOLNAME} is used in replay mode on Bob's side.}\delete{
Bob uses \texttt{\TOOLNAME} in replay mode.} It first determines an efficient {\em replay sequence} or replay order, i.e., a plan for execution that includes checkpoint caching decisions. To do so \texttt{\TOOLNAME} \revision{inputs} a cache size bound, $B$, and then executes a heuristic algorithm on the execution tree received from Alice to determine the most cost efficient replay sequence for that cache size. Computing a cost-optimal replay sequence for \revision{multiple versions of a program}\delete{a given multiversion program} with a cache bound is an NP-hard problem as we show in Section~\ref{sec:problem} and so we describe some efficient heuristics for this purpose (Section~\ref{sec:solution}). Finally, once the replay sequence is computed, \texttt{\TOOLNAME} uses this replay sequence to \textit{compute, checkpoint, restore-switch} REPL program cells or \textit{evict} stored checkpoints from cache.

\paragraph*{Our assumptions}

Our basic assumption is that Bob wishes to independently verify the results from Alice's versions but is time constrained to repeat all her versions. \revision{We do not make any assumptions on the types of edits that differentiates one version from the next. Thus, Alice can change values of parameters, specifications of datasets, models, or learning algorithms. She can also add or delete entire cells. We illustrate possible changes via red boxes (Figure~\ref{fig:notebook}) across program versions. We only assume that edits result in valid executions, which do not terminate in an error, and, each version is executed in the natural order, top to bottom.}
\delete{Apart from this we make some other mild assumptions about Alice and Bob: We assume Alice has 
developed code in paragraphs, which is typically how developers work. We also assume Alice makes edits to create versions such that (i) edits are on the same REPL program $P$ with a starting specification that is edited, and further edited, (ii) edits are atomic and result in valid executions, which do not terminate in an error, and finally
(iii) edits preserve natural ordering before the edited cell.

\texttt{\TOOLNAME} works with REPL cells, but Alice as a user of \texttt{\TOOLNAME} is not constrained to program in REPL style. If the code is not in REPL style \texttt{\TOOLNAME} translates it into cells of a REPL program during preprocessing. This preprocessing takes care to not split functions or control flows into separate cells. Thus every input program is automatically transformed into an equivalent REPL program and then entered into the \texttt{\TOOLNAME} pipeline.}

\section{The Multiversion Replay Problem}
\label{sec:problem}

\begin{figure}
\small
    \centering
    \begin{tabular}{|l|l|}
        \hline
        $L_j$ & REPL program version\\
        $x_i$ & Cell in $L_j$\\ 
        $h_i$ & Hash of source code in $x_i$ \\
        $g_i$ & Cumulative hash of source code and ext. dependencies till $x_i$\\
        $ps_i$ & Program state at end of $x_i$  \\
        $sz_i$ & Size of $ps_i$ \\
        $\delta_i$ & Computation time to reach $ps_i$ from $ps_{i-1}$ \\
        $T$ & Execution tree combining overlapping $L_j$'s\\
        $B$ & Fixed cache size\\
        $R$ & Replay sequence for $T$\\
        $O_t$ & Operation at step $t$ in $R$\\
        $S_t$ & Set of program states in cache after step $t$ in $R$\\
        $\delta(R)$ & Computation time for $R$\\
        \hline
    \end{tabular}
    \caption{Notation used in Section~\ref{sec:prelim} and Section~\ref{sec:problem}} 
    \label{fig:notation}
\end{figure}

\revision{We now describe the multiversion replay problem. Figure~\ref{fig:notation} summarizes the symbols used in Section~\ref{sec:prelim}. In the replay mode, \texttt{\TOOLNAME} inputs an execution tree, $T$, and a fixed cache size, $B$, to solve the multiversion replay problem.} 
\revision{We define the execution tree as:}
\begin{definition} \label{def:extree}(Execution Tree)
 An execution tree $T=(V,E)$ is a tree in which \revision{each program state is mapped to a node and reusable program states}\delete{equal} \delete{cells} across the \delete{enhanced specifications of }different versions are mapped to the same node. Each root to leaf path in $T$ corresponds to a distinct \revision{version $L_i$.}\delete{$\overline{L} \in \overline{\LL}$\revision{, in which $|\overline{L_i}| \neq |\overline{L_j}|$.}} 
 \end{definition}

\begin{figure}[t]
    \centering
    \includegraphics[width=0.8\linewidth]{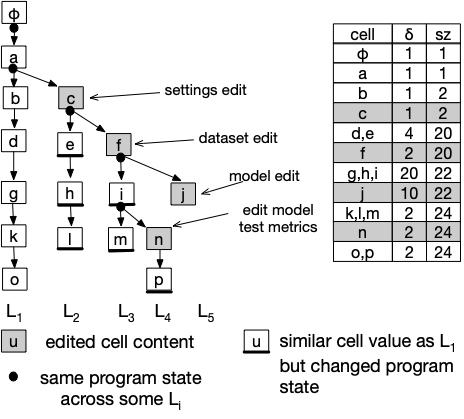}
    \caption{\revision{The enhanced specifications  of versions (without lineages for simplicity) represented as a tree. The cell $i$ appears similar to $h$ but has a changed program state due to edited $f$. Both  $m$ and $n$ proceed from $i$'s state.}}
    \label{fig:tree}
\end{figure}

 \noindent {\bf Example.}
 Figure~\ref{fig:tree} shows the execution tree created from \delete{enhanced specifications of} five versions. 
 In this tree, each root to leaf path corresponds to \revision{version}\delete{an enhanced specification $\overline{L}_i$ of a version} \revision{$L_i$}\delete{$i$}. In $L_1$ there is an edit to settings of the program at cell $b$, resulting in $L_2$ and a branch at $a$, the last common node across $L_1$ and $L_2$. Similarly, in $L_3$ there is a dataset change to $L_2$ at cell $e$, resulting in $L_3$ and a branch at $c$, the last common node across $L_2$ and $L_3$. 
 The common nodes till a branch in the tree correspond to the subsequence of cells that are \revision{reusable}\delete{equal} across \revision{versions}\delete{specifications}. The  tree  branches  at  a  cell node, subsequent  to  which cells are \revision{not reusable}\delete{similar but not equal}. \texttt{\TOOLNAME} computes cell \revision{reusability}\delete{equality} using execution lineages.  We will discuss how this is done via system calls in detail in Section~\ref{sec:chex}.

The multiversion replay problem is an optimization problem that arises when multiple versions of a program, each previously executed, are replayed as a collection. Once the multiversion program is represented as an execution tree it is clear that there is some advantage in not replaying the common prefixes of this tree. 

\begin{figure}[t]
    \centering
    \includegraphics[width=\linewidth]{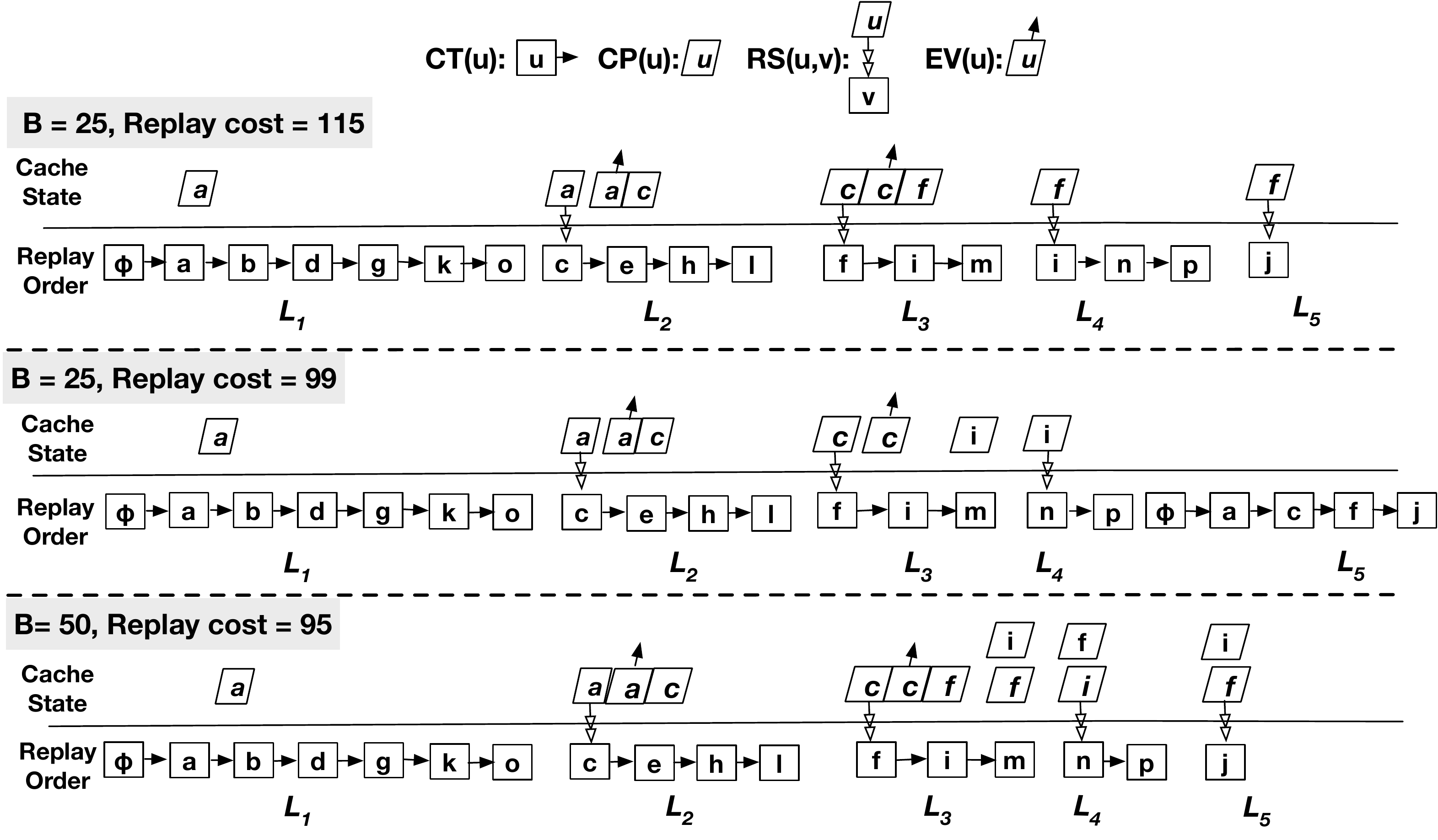}
    \caption{Replay sequences for the execution tree in Figure~\ref{fig:tree} showing use of operations and the state of the cache. }
    \label{fig:replaysequenceexamples}
\end{figure}

\noindent {\bf Example.}  If we replay the five versions of Figure~\ref{fig:tree} sequentially we incur total cost of 129. On the other hand, assuming a cache size of 25,
if we store the \textit{checkpoint} at common prefixes, \textit{restore-switch} the checkpoint later to \revision{avoid computing the common prefix for the next version}, and \textit{evict} the previous checkpoint to store a new one, the replay cost is reduced to 114 as shown in the first replay sequence of   Figure~\ref{fig:replaysequenceexamples}. In the second figure we see that a different set of checkpointing decisions can improve the cost even when the cache size remains the same. Finally we see that increasing the space to 50 further improves replay costs to 95. 

Under these operations, and given an execution tree and a fixed  amount  of  space  for  storing  checkpoints, the multiversion replay problem aims to  determine  a  replay  sequence  that  has  the  minimum  replay costs.  We define a general replay sequence as follows:

\begin{definition}[Replay sequence]
Given execution tree $T = (V,E)$ and  a  cache  of  size $B$, a replay sequence $R$ consists of $m$ steps such that step $t$  specifies the operation $O_t$ performed and the resulting state of the checkpoint cache $S_t$, \textit{i.e., }
\[ R = [(O_t,S_t): 0 \leq t \leq m]\] 
We will use the term replay order interchangeably with the term replay sequence.
\end{definition}

At the initial step $S_0$ is empty and the root of the tree is computed. At any given step $t$, $O_t$ is of one of the following four types.  Here, $u_j$ and $u_k$ are nodes in $V$, the vertices of $T$.
\begin{itemize}
\setlength{\itemsep}{1pt}
\item Compute $CT(u_j)$: computes $u_j$;
\item Checkpoint $CP(u_j)$: checkpoints $u_j$ into the cache;
\item Restore $RS(u_j,u_k)$: restores a previous checkpointed $u_j$ in cache and switches to $u_k$ where $u_j = parent(u_k)$; and 
\item Evict $EV(u_j)$: evicts a previous checkpoint $u_j$ from cache; 
\end{itemize}

The cache size can never exceed $B$, i.e. $|S_t| \leq B$ for $0 \leq t \leq m$.  Further, an operation $O_t$ at step $t$ can only be performed on $u_j, u_k$ under the following constraints:

\begin{itemize}
\setlength{\itemsep}{1pt}
    \item \textit{Checkpoint from working memory:} A node in the execution tree is checkpointed only if it was computed in some previous step, after which there are only some evictions (to make space), if at all,  \textit{i.e.,} if $O_t = CP(u_j)$ $\implies$ $S_t = S_{t-1} \cup \{u_j\}$ and $O_{t-i} = CT(u_j)$, for some $1\leq i \leq t$, and $
    O_{t'} = EV(u_{t'})$ for $t-i < t' < t$. 
    \item \textit{Restore from cache and switch to child:} A node is restored only if it was in cache in a previous step, and without altering cache state, switches to one of its children in the execution tree, which is computed next \textit{i.e.,} if $O_t = RS(u_j,u_k)$ $\implies$ $u_j \in  S_{t-1}$, $S_t = S_{t-1}$, $O_{t+1} = CT(u_k)$.
    \item \textit{Evict from cache:} A node is evicted from cache and alters its state, \textit{i.e.,} if $O_t = EV(u_j)$  $\implies$ $u_j \in  S_{t-1}$, $S_t = S_{t-1}-\{u_j\}$.
    \item \textit{Continue computation:} Continue computing a node if its parent was being computed or if its parent was restored, \textit{i.e.,} if $O_t = CT(u_j)$  $\implies$ $O_{t-1} = CT(parent(u_j))$ or $O_{t-1} = RS(parent(u_j), u_j)$ and $S_t = S_{t-1}$ \revision{or t= 1 and $u_j$=root of the tree $T$.}
\end{itemize}

We assume the operations 
generate \textit{complete} and \textit{minimal} sequences. A replay sequence is complete if all leaf nodes of the tree $T$ appear in $R$, and is 
minimal if no $u_j$ that is in cache is recomputed. 
\begin{problem}[The Multiversion Replay Problem (\probname)]
  \label{prb:offline-problem}
Given tree $T(V,E)$, the multiversion replay problem is to find a complete replay sequence $R$ that minimizes
\[ \ccost(R) = \sum_{i=0}^{|R| = m} \delta_{O_t},\] in which   $\delta_{O_t} = \delta_j$, when $O_t = CT(u_j)$, and $\delta_{O_t} = 0$ otherwise.
\end{problem}

In \probname, we assume the cost of checkpoint, restore-switch, and evict operations to be negligible. Thus, the only cost considered is the cost of computing the cells. 
Determining  the  minimum  cost  replay  order  leads  to  a natural   trade-off   between   computational   cost   of   cells  and fixed-size cache storage occupied by the checkpointed state of the cells. Thus, to optimally  utilize a  given  amount  of  storage  we must determine  for each cell whether its next cell be recomputed, or some other cell be recomputed by checkpointing the state of the current cell. We state that determining the replay order is computationally hard.

\begin{theorem}
\label{thm:np-hard}
  \probname\ is NP-hard.
\end{theorem}

\ignore{
\revision{We show that the decision version of \probname is NP-hard. Given an execution tree $T$, a cache size parameter $B > 0$ and a total cost parameter $\totcost > 0$, define $\decprob(T, B, \totcost)$ to be the decision problem with answer YES if there is a replay sequence of $T$ with cost at most $\totcost$ and size of cache at most $B$, and with answer NO otherwise.

The proof is by reduction from the decision version of bin packing. In outline the proof works by constructing an execution tree whose depth 1 nodes have checkpoint sizes corresponding to the size of the items to be packed into bins in the bin packing problem. The $B$ of $\decprob(T, B, \totcost)$ is set to the size of the bins. In order to force caching we keep $\totcost$ small and add nodes below the depth 1 nodes so that each of the level one nodes has to be cached when first computed. We are able to show that by carefully adding subtrees below the depth 1 nodes we are able to prove a tight relationship between the two problems, i.e., the bin packing decision problem gives a Yes answer iff $\decprob(T, B, \totcost)$ gives a yes answer.
}

We omit the proof due to space restrictions, referring the reader to the full version of this paper available at~\cite{CHEX-techreport}.
}


To prove this first define the decision version of \probname. Given an execution tree $T$, a cache size parameter $B > 0$ and a total cost parameter $\totcost > 0$, define $\decprob(T, B, \totcost)$ to be the decision problem with answer YES if there is a replay sequence of $T$ with cost at most $\totcost$ and size of cache at most $B$, and with answer NO otherwise.

\begin{figure}[h]
    \centering
    \includegraphics[width=1.0\linewidth]{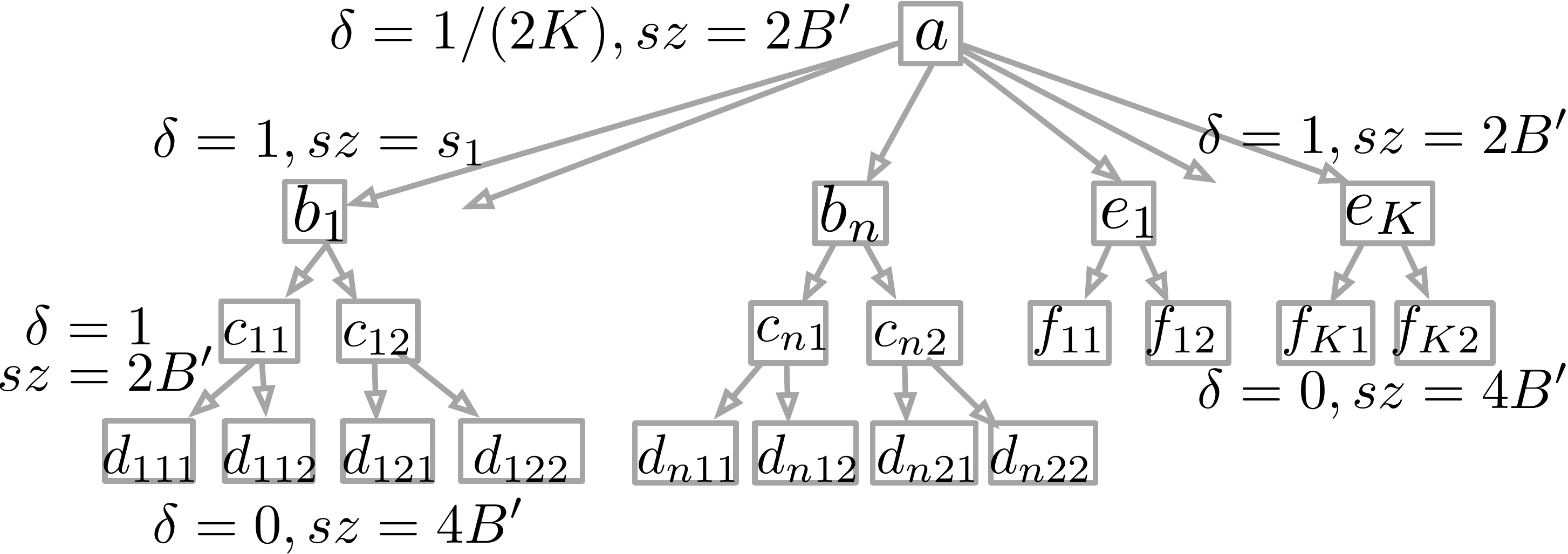}
    \caption{The ``gadget" for reducing bin packing to \probname.}
    \label{fig:proof}
\end{figure}

\begin{proof}

The proof is by reducing bin packing \cite{kleinbergnpc} to $\decprob$. For $n > 0$; given a bin size $B' > 0$; a set $A$ of $n$ items with sizes $s_1, \ldots, s_n$; such that $s_i \leq B'$ for $1 \leq i \leq n$; and an integer parameter $K > 0$, define $BP(A,B',K)$ to be the problem with answer YES if the items in $A$ can be packed into at most $K$ bins, and NO otherwise.  (Assume $K < n$ otherwise the problem is trivial.)

Given any instance $BP(A,B',K)$ we next show how to construct a corresponding instance of $\decprob(T, B, \totcost)$ which returns YES if and only if $BP(A,B',K)$ returns YES.  Further this construction is made in time polynomial in the size of the input to $BP(A,B',K)$, proving that $\decprob$ is at least as hard as bin packing, i.e., it is NP-hard.

In the corresponding instance $\decprob(T, B, \totcost)$, $B$ is set to $3B'$ and $\totcost$ to $(3n + K + 1/2)$.  The execution tree $T$ has a root node $a$ with $(n + K)$ child subtrees, as described below (see Figure \ref{fig:proof}).  Also $\delta_a = 1/(2K)$ and $sz_a = 2B'$.

The first $n$ child subtrees correspond to the $n$ items in $A$.  In the $i$th such subtree, the root node is $b_i$ with cost $\delta = 1$ and size $sz = s_i$---the size of the $i$th item in $A$.  $b_i$ is a child of $a$.  The rest of the child subtree is the same irrespective of $i$: $b_i$ has two child nodes $c_{i1}$ and $c_{i2}$, each with cost $\delta = 1$ and size $sz = 2B'$.  Further $c_{i1}$ has two child nodes $d_{i11}$ and $d_{i12}$, each with cost $\delta = 0$ and size $sz = 4B'$.  Similarly $c_{i2}$ has two child nodes $d_{i21}$ and $d_{i22}$, each with cost $\delta = 0$ and size $sz = 4B'$.  Observe that when any of $c_{i1}$ or $c_{i2}$ is cached, the root $a$ cannot be in the cache.  This will be used later to restrict when $\decprob(T, B, \totcost)$ can return YES.

The remaining $K$ child subtrees of $a$ are identical.  In the $j$th such subtree, the root node is $e_j$ with cost $\delta = 1$ and size $sz = 2B'$.  $e_j$ is a child of $a$.  $e_j$ itself has two child nodes, $f_{j1}$ and $f_{j2}$, each with $\delta = 0$ and size $sz = 4B'$.

Let $\decprob(T, 3B', (3n + K + 1/2))$ return YES, and let $R$ be a corresponding satisfying replay sequence.  Now $T$ has $(3n + K)$ nodes with cost $\delta = 1$.  These are the $n$ nodes each of type $b_i$, $c_{i1}$ and $c_{i2}$, and the $K$ nodes of type $e_j$.  All other nodes in $T$ have cost $\delta = 0$, except for $a$, which has cost $1/(2K)$.  Since each node in $T$ is computed at least once, the YES answer is only possible if each of the nodes with cost $\delta = 1$ is computed exactly once and $a$ is computed at most $K$ times.  This in turn has the following implication.  Each node with cost 1 (i.e. of types $b_i, c_{i1}, c_{i2}$, and $e_j$) have two children.  If they are computed exactly once, they have to be checkpointed and cached as soon as they are computed, else the second child would require a recomputation.

Divide the replay sequence $R$ into phases, each beginning with a recomputation of the root $a$.  Since $a$ is computed at most $K$ times, there are at most $K$ phases.  Whenever an $e_j$ is computed, it is cached, which requires $a$ to be evicted from the cache if it was in it, to make space.  This implies that each node of type $e_j$ is computed and cached in a different phase.  Further, any $e_j$ in a phase has to be the last child of $a$ to be computed in a phase; any nodes of type $b_i$ have to be computed and cached before the $e_j$.  For a similar reason none of the nodes of type $c_{i1}$ or $c_{i2}$ can be computed and cached before an node of type $e_j$ in a phase.

The above arguments imply that $R$ has exactly $K$ phases, and in each phase some $b_i$-type nodes are computed and immediately cached, followed by the computation and caching of a node of type $e_j$.  Following this, nodes of type $c, d$, and $f$ are computed and/or cached in some order---the details of which we can ignore.  More importantly, when the last $b_i$ node is computed and cached and before the $e_j$ node is computed, node $a$ has to be in cache. At this point the sum of the sizes of the $b_i$ nodes cached in this phase is at most $3B' - sz_a = B'$.  Thus the $K$ phases give us a partition of the items in $A$ into $K$ bins where the items in bin $k$ correspond to the $b_i$ nodes computed in phase $k$, for $1 \leq k \leq K$, and whose sizes have to add up to at most $B'$.  This implies that $BP(A, B', K)$ has a YES answer.

Reversing the argument to show that a YES answer to $BP(A, B', K)$ implies a YES answer to $\decprob(T, 3B', (3n + K + 1/2))$ is straightforward.  Simply construct $K$ phases corresponding to the $K$ bins used (some bins may be empty), and in each phase compute the nodes in the order described above.

Finally, observe that the input to $\decprob(T, 3B', (3n + K + 1/2))$ is $O(7n + 3K)$ which is clearly polynomial in the size of the input to $BP(A, B', K)$.
\end{proof}

\section{Heuristic Solutions}
\label{sec:solution}


From Theorem~\ref{thm:np-hard} we know that it is unlikely we will find polynomial time solution to \probname. Accordingly, we present two efficient heuristics for this problem. 
Both heuristics restrict our exploration of the search space to solutions in which the execution order of the nodes of the execution tree corresponds to a DFS traversal of the tree---a natural, simple order in which to approach the replay of the tree.  In order to formalize this notion we present some definitions.  In the following, for sake of brevity, we specify only the compute $CT(u_j)$ type operations in replay sequences.  The other operations (checkpoint, restore, evict) are separately specified. In this briefer format, each step of a replay sequence is of the form $(u_t, S_t)$ specifying that at step $t$, $u_t$ is computed, and the resulting cache is 
$S_t$.

\begin{definition}[Ex-Ancestor replay sequence]
\label{def:exA}  
  Suppose  $T = (V,E)$ is an execution tree. Given any replay sequence $R = \{(u_t,S_t): 1 \leq t \leq T\}$ we define its {\em first appearance order} to be $i_1 < i_2 < \cdots < i_{|V|}$ such that $u_{i_j}$ is the first appearance of a node $v_j$ of T in $R$. We call the indices $i_1, \ldots, i_{|V|}$ as {\em first appearances} and all other indices as {\em repeat appearances}.
  For a replay sequence $R$, for each $j \in \{2, \ldots, |V|\}$ the sequence of cells $u_{i_{j-1}+1}\ldots u_{i_j-1}$ is called the {\em helper sequence} for $v_{j}$. If the helper sequence for $v_j$ forms a path from an ancestor of $v_j$ to $v_j$ for each $j$ then $R$ is called an {\em ex-ancestor replay sequence}.
\end{definition}
We observe that in an ex-ancestor replay sequence if the helper sequence of $v_j$  is non-empty then it either begins with the root of $T$ or with a node whose parent is in $S_{i_{j-1}}$.

\revision{We illustrate this definition with an example. 
  Consider the tree in Figure~\ref{fig:tree}. Assume for now that cache size $B = 0$ and consider the following replay sequence: $$a, b, d, g, k, o, \bba, c, e, h, l, \bba, \bbc, f, i, m, \bba, \bbc, \bbf, \bbi, n, p, \bba, \bbc, \bbf, j$$ where bold font indicates repeat appearance nodes. Here the indices 1, 2, 3, 4, 5, 6, 8, 9, 10, 11, 14, 15, 16, 21, 22, and 26  are first appearances and all others are repeat appearances. Let's take the example of node $n$. It's first appearance index is $21$. Its helper sequence extends from indices 17 to 20 and contains $\bba, \bbc, \bbf, \bbi$. This is a simple path in the tree beginning from the node containing $a$ which is an ancestor of $n$. In fact it is easy to verify that this sequence is an ex-ancestor replay sequence.

The question arises: Are there meaningful replay sequences that are not ex-ancestor replay sequences? For example, would it make sense to modify $n$'s helper sequence and make it $\bba, \bbc, e, \bbf, \bbi.$ It appears that the extra computation of $e$ is superfluous and so a priori it is not obvious that such replay sequences are meaningful from the point of view of efficient replay. Therefore we focus on ex-ancestor replay sequences. We conjecture that an optimal solution to \probname\ will be such a sequence. 
}
\begin{definition}[DFS-based replay sequence]
  Suppose  $T = (V,E)$ is an execution tree for a collection of traces $\CC$. We say a complete and minimal replay sequence $R$ is a {\em DFS-based replay sequence} if $R$ is an ex-ancestor sequence and the first appearance order of $R$ is a DFS-traversal order of $T$. 
 \end{definition}
 
 
 
\ignore{
A DFS-based replay sequence is essentially a sequence in which the nodes of the execution tree $T$ are processed in a manner corresponding to a DFS-order.  This does not imply that the replay sequence itself is a DFS-traversal of $T$, but the first-appearance subsequence of the replay sequence is, as we shall see.
 
Consider the tree in Figure \ref{fig:tree}.  Suppose we wanted to replay the nodes of the tree in a manner corresponding to the following DFS-traversal: $a, b, d, g, k, c, e, h, l, f, i, m, n, j, o$.

We illustrate the definition with an example.   
Assume for now that cache size $B = 0$ and consider the following replay sequence: $$a, b, d, g, k, \bba, c, e, h, l, \bba, \bbc, f, i, m, \bba, \bbc, \bbf, \bbi, n, \bba, \bbc, \bbf, j, o,$$ where bold font indicates repeat appearance nodes.  It is easy to verify that this is an ex-ancestor replay sequence}

Note that first appearance sequence of the example discussed below Definition~\ref{def:exA} gives us a DFS-traversal of $T$. Hence this is a DFS-based replay sequence for the tree of Figure~\ref{fig:tree}.

Now assume a cache size $B = 25$ and the caching decisions made according to the first replay sequence in Figure \ref{fig:replaysequenceexamples}.\footnote{All three replay sequences in Figure \ref{fig:replaysequenceexamples} are DFS-based.}  The corresponding replay sequence is similarly: $a, b, d, g, k, o, c, e, h, l, f, i, m, \bbi, n, p, j$.  Since $a, c,$ and $f$ are cached at appropriate junctures, the only node with a non-empty helper sequence is $n$ and the length of this sequence is just one, i.e., there is only one cell that has to be recomputed apart from its first appearance computation.  For the second and third sequence in Figure \ref{fig:replaysequenceexamples} the number of recomputations are similarly three ($\bba, \bbc, \bbf$) and zero respectively.

We are able to explicitly bound the number of DFS-based replay sequences.
\begin{proposition}
\label{prp:no-dfs-based}
  Suppose  $T = (V,E)$ is an execution tree for a collection of traces $\CC$ such that $|V| = n$ and the height of $T$ is $h$. Let $b_u$ be the number of children of node $u \in V$ and let
  \[\bar{b} := \frac{1}{n}\sum_{u \in V} b_u \log b_u.\]
Then, the number of DFS-based replay sequences of $T$ is $O(2^{n(h+\log h+\bar{b})})$.
\end{proposition}
\begin{proof}
  Let us fix a DFS traversal order. The helper sequence preceding (and including) each node can be at most $h$ in length and hence the length of a replay sequence can be no longer than $hn$. Since each helper sequence is an ex ancestor path to a node of the tree, we can have at most $h$ choices of a helper sequence at each node. Therefore there are at most $h^n$ different sequences that can qualify to be DFS-based replay sequences.  Note that at each point of one of these sequences of cells we can decide to either cache the cell that we have just computed (this may require the eviction of something previously cached) or to not cache it. Hence there are at most $2^{nh}$ replay sequences associated with each of the $h^n$ different sequences that we got for a single DFS traversal order. 
  This gives us an upper bound.

  To compute the number of DFS traversal we simply permute the children visited by DFS at each step to get $\prod_{u \in V} b_u!$ which can be rewritten as $2^{n\bar{b}}$ by using Stirling's approximation. Multiplying we get the result.
\end{proof}  
Since $h$ is $\Omega(\log n)$ from Proposition~\ref{prp:no-dfs-based} it appears that the space of possible solution is superexponential. In order to control the complexity of our solutions we restrict the solution space in two different ways and define two heuristics. 

\ignore{
\begin{itemize}
\item {\em Persistent Root Policy Greedy (PRP)}. We restrict ourselves to solutions in which if a node $u$ is ever in cache, it is required to be in cache until the entire subtree rooted at $u$ is computed.  This makes the particular DFS order irrelevant, and reduces the upper bound of possible solutions to $O(2^{n(h + \log h)})$. Our heuristic constructs a solution greedily in this space.
A detailed discussion is presented in Section~\ref{sec:heuristic:greedy}.

\item {\em \bcalgo\ (PC)}. We (a) work with an arbitrary fixed DFS order and (b) allow a cell to be cached only if it is the parent of the node whose first computation is forthcoming, and (c) don't allow a cached node to be evicted till we have executed all the leaves of the subtree of the child for which it has been cached.  Since node $u\in V$ is a candidate for caching at each of its children, the number of possible solutions, given that we work with a fixed DFS order, is reduced to $2^{n\bar{b}}$. We present the details of this heuristic in Section~\ref{sec:heuristic:parent}.
\end{itemize}
}

\subsection{Persistent Root Policy Greedy Algorithm}
\label{sec:heuristic:greedy}

Within the space of DFS-based replay sequences, for our first heuristic, we propose the following caching policy: {\em A cell can be cached only when it is first computed. Once cached the cell remains in the cache till every leaf of the subtree rooted at the node containing cell is computed.} We call this the {\em DFS Persistent Root policy}.

Given a DFS traversal order this policy reduces the size of the solution space to $O(2^n)$ which is still exponential in the size of the tree. We present a greedy algorithm called {\em Persistent Root Policy Greedy} ({\bf PRP}) that helps find a good solution in polynomial time.
\begin{algorithm}[t]
\caption{A greedy algorithm that takes as input execution tree $\CT,$ and a cache size parameter $B$. It outputs list $S$ of nodes to be cached under the DFS Persistent Root policy.}
\label{alg:greedy}
\begin{algorithmic}[1]  
\Function{{\bf PRP}}{$\CT,B$}
  \State $S \gets \emptyset$
  \State $f \gets \text{True}$ \Comment{$f$ is True while greedy is able to extend its solution }
  \State $r \gets \text{root}(\CT)$
  \State Set $\min \gets$ \Call{DFSCost}{$r,S,B,0$} \Comment{The function gives us the cost of a DFS-based replay sequence for $\CT$ given a list of nodes $S$ that must be cached when first computed.}

  \While{$f$ is True and $S \ne V$}
    \State $f \gets \text{False}$
    \For {each $u \in V \setminus S$}
    \If {\Call{DFSCost}{$r,S\cup\{u\},B,0$} $< \min$}
    \State $f \gets \text{True}$ \Comment{We can extend the solution}
    \State $u^* \gets u$ \Comment{$u*$ is the current best candidate}
    \EndIf
    \EndFor
    \If{$f$ is True}
    \State $S \gets S \cup \{u^*\}$
    \State $\min \gets$ \Call{DFSCost}{$r,S,B,0$}
    \EndIf
    \EndWhile
\State {\bf return} $S$    
\EndFunction
\end{algorithmic}
\begin{algorithmic}[1]
\Function{DFSCost}{$u,S,B,b$} \Comment{$u$ is a node of $\CT$; $b$ is the cache budget used by the path from the root of $\CT$ to $u$. Called with $u = \text{root}(\CT)$ and $b=0$ this returns the cost of computing the entire tree.}
  \If{$u \in S$ and $b + \storage{u}>B$}
  \State {\bf return} $\infty$ \Comment{Cache size infeasibility detected}
  \EndIf
  \State $c_u \gets \text{cost of computing $u$ from nearest ancestor in $S$}$
  \If{$u$ has no children}
  \State {\bf return} $c_u$ 
  \EndIf
  \State $\text{sum} \gets 0$ 
  \For{each $v$ that is a child of $u$}
  \If {$u\in S$}
  \State $\text{sum} \gets$ \Call{DFSCost}{$v,S,B,b+\storage{u}$}
  \Else
  \State $\text{sum} \gets$ \Call{DFSCost}{$v,S,B,b$} + $c_u$ \Comment{$u$ is not cached so must be recomputed for each child}
  \EndIf
  \EndFor
  \If {$u \in S$}
  \State $\text{sum} \gets \text{sum} + c_u$ \Comment{$u$ must be computed once}
  \EndIf
  \State {\bf return} $\text{sum}$
\EndFunction
\end{algorithmic}  
\end{algorithm}
We present the listing of {\bf PRP} as Alg.~\ref{alg:greedy}. The algorithm begins with the baseline cost (stored in $\min$) of a DFS-based replay sequence in which no node is cached and seeks out the node of the tree whose addition to the list $S$ achieves the maximum improvement over the baseline. This process continues incrementally while it is possible to include another node in the list. The process will stop when the subroutine {\em DFSCost} tells us that there is no node remaining in $V \setminus S$ that can be included in $S$. Typically this will happen because for every node $u$ remaining in $V \setminus S$, the cache will be full when it is encountered in the DFS order. In such a situation {\em DFSCost} will return $\infty$.
This algorithm takes $\theta(n^2)$ time to find each candidate to include in the list of nodes to be cached, and there are potentially $O(n)$ such nodes. Therefore the time complexity of this algorithm is $O(n^3)$. However there is no guarantee of optimality.

{\bf PRP} is a greedy algorithm that seeks, at each iteration, to pick for caching the vertex of the execution tree that minimizes the cost. However, it can be easily modified to choose a vertex that minimizes the cost incurred {\em per unit of cache memory consumed.} Normalizing by size is a common measure for object caches~\cite{cao1997cost,malik2005bypass}. We experimentally study both these variants in Section~\ref{sec:experiments}. We will refer to the cost-minimizing version as {\bf PRP-v1} and the ratio minimizing version as {\bf PRP-v2}.

\subsection{\bcalgo\ Algorithm}
\label{sec:heuristic:parent}
We now present a second heuristic that, while still not being optimal, searches a superset of the portion of the solution space searched by {\bf PRP}. For each $u \in V$ it seeks to partition the children of $u$ into two sets: $P_u$ of nodes for which it is better to cache $u$ for the computation of the corresponding child subtrees, and $\overline{P}_u$ for which it is not. As in Persistent Greedy, caching choices once made persist here as well.

\ignore{


Suppose we wanted to search the space of all DFS-based replay sequences to find one with optimal cost.  Roughly speaking, it could be done in the following recursive manner: at each node consider the replay cost of each permutation of the child subtrees, and for each permutation consider all possible valid cache contents between every pair of consecutive child subtree computations.

The above would be prohibitively expensive.  The \bcalgo\ {\bf (PC)} algorithm takes a simpler approach.
}

\ignore{


For each $u \in V$ it seeks to partition the children of $u$ into two sets: $P_u$ of nodes for which $u$ is to be cached before the node is first encountered and $\overline{P}_u$ for which $u$ is not to be cached. As in Persistent Greedy, caching choices once made persist here as well.
}

The listing of the essential recursive \bcalgo\ is presented as Alg.~\ref{alg:bcalgo}.  When called with $(u, S)$ we explore the situation in which we are given the set $S$ of ancestors of $u$ that will be in cache while the subtree rooted at $u$ is computed.
In case $u$ happens to be a leaf, no further decisions are needed, and we simply return the cost of computing $u$ given cache $S$ (Lines~\ref{ln:leaf-i}-\ref{ln:leaf-e}).  Else, we need to determine what is best for each child $u_i$ of $u$: Should the subtree rooted at $u_i$ be computed with $S $ as is, or is it better to augment the cache with $u$ (denoted $\bSu$).  In the former the subtree may be forced to recompute $u$ multiple times, in the latter cache space which may be useful down the subtree is used up.  The two costs are computed recursively (Lines~\ref{ln:with-u} and~\ref{ln:without-u}), and the child is assigned to the set $P_u$ or $\overline{P}_u$ corresponding to the lower cost (Lines~\ref{ln:partition}-\ref{ln:partition-end})).  Note that when adding $u$ to the cache is infeasible, i.e. $|\bSu| > B$, we make the first choice for each  node, i.e. assign them all to $P_u$. (Lines~\ref{ln:no-space-left}-\ref{ln:no-space-left-end}).  Finally, we return the cost value up the recursion stack (Line~\ref{ln:return-value}).

\ignore{


The listing of \bcalgo\ is presented as Alg.~\ref{alg:bcalgo}. Note that in Lines~\ref{ln:with-u} and~\ref{ln:without-u} \ignore{there is a branch and} two  recursive calls are made. Effectively that means that when the main function is called at node $u_i$ in Lines~\ref{ln:with-u} and~\ref{ln:without-u}, a partial decision regarding the caching of the parents of all the nodes up to {\em and including} $u$ (except the root) has already been made and the function call is expected to extend the decision to the child $u_i$ of $u$. If the children of $u$ are leaves, no further caching is needed and we simply return the cost (Lines~\ref{ln:leaf-i}-\ref{ln:leaf-e}). If $u$ is an internal node, we need to branch: whether we proceed with the cache received in the function call or we augment the cache received in the function call with $u$. Note that this augmentation may not be feasible in which case we simply put all of $u$'s children in $\bar{P}_u$ (Lines~\ref{ln:no-space-left}-\ref{ln:no-space-left-end}). If the augmentation is feasible we branch (Lines~\ref{ln:with-u}-\ref{ln:without-u}). Depending on the cost values obtained from the recursive calls we assign children to $P_u$ or $\bar{P}_u$ (Lines~\ref{ln:partition}-\ref{ln:partition-end}). Finally, we return the cost value up the recursion stack (Line~\ref{ln:return-value}).

}

The essential recursive algorithm {\bf PC} needs to be implemented using standard dynamic programming memoization and backpointers (see, e.g., \cite{clrsdynaprog}). Once a call with input $(u, S)$ is complete, the corresponding return cost value and the two sets $P_u$ and $\overline{P}_u$ are recorded.  The initial call is with $(\text{root}(\CT), \emptyset)$.  This returns the cost of the optimal replay sequence for the entire $\CT$.  To construct the replay sequence itself, ``follow the backpointers'': Start with $u = \text{root}(\CT)$ and $S = \emptyset$.  If for the corresponding call, $P_u$ is not empty, compute $u$ (possibly by restoring the closest ancestor in the current cache) and checkpoint it.  Update $S$ to include $u$.  Then recursively compute the subtrees rooted at the nodes in $P_u$.  Next update $S$ to remove $u$. Following this, recursively compute the subtrees rooted at the nodes in $\overline{P}_u$, if any.

The above implementation takes time and space proportional to the total number of child nodes encountered over all recursive calls.  For each $u \in V$, at most one recursive call is made for each possible set of ancestors in the cache.  The number of different ancestor sets is at most $2^h$.  Thus the total time taken is $O(2^{h}\sum_{u \in V} b_u)$.

\begin{algorithm}[t]
\caption{A recursive algorithm the computes for a tree rooted at $u$ the lowest DFS-based replay cost for a given cache $\bSin$.  The child subtrees of $u$ are allowed to choose between executing with $\bSin$ or in addition caching $u$.}
\label{alg:bcalgo}
\begin{algorithmic}[1]  
\Function{\bcalgopc}{$u, \bSin$}
  \If{$u$ is a leaf} \label{ln:leaf-i}
     \State $a \gets \text{nearest ancestor of $u$ in $\bSin$}$
     \Return{\text{cost of computing $u$ from $a$}} \label{ln:leaf-e}
     \State {\Comment{If $a$ doesn't exist, return cost of computing $u$ from scratch.}}
  \EndIf
  \State $\bSu \gets \bSin \cup \{u\}$ \Comment{$\bSu$ is cache that also includes $u$.}
  \If{size of cache $\bSu > B$} \label{ln:no-space-left}
  \State {\Comment{Caching $u$ is not a option; process its children with $\bSin$.}}
  \State $P_u \leftarrow \emptyset$; $\overline{P}_u \leftarrow \children(u)$.
    \Return $\sum_{u_i \in \overline{P}_u} \Call{\bcalgopc}{u_i, \bSin}$
  \EndIf \label{ln:no-space-left-end}
  \State $P_u \gets \emptyset, \overline{P}_u \gets \emptyset$
  \State{\Comment{$P_u$ will collect the nodes for which caching parent $u$ is cheaper.}}
  \For {each $u_i \in \children(u)$}
     \State $\bccost(u_i, \bSu) \gets \Call{\bcalgopc}{u_i, \bSu}$ \label{ln:with-u}
     \State $\bccost(u_i, \bSin) \gets \Call{\bcalgopc}{u_i, \bSin}$ \label{ln:without-u}
     \If{$\bccost(u_i, \bSu) \leq \bccost(u_i, \bSin)$} \label{ln:partition}
        \State $P_u \gets P_u \cup \{u_i\}$
     \Else
        \State $\overline{P}_u \gets \overline{P}_u \cup \{u_i\}$ \label{ln:partition-end}
     \EndIf
  \EndFor
  \Return{$\sum_{u_i \in P_u} \bccost(u_i, \bSu) + \sum_{u_i \in \overline{P}_u} \bccost(u_i, \bSin)$} \label{ln:return-value}
\EndFunction
\end{algorithmic}
\end{algorithm}

\section{Execution Tree}
\label{sec:chex}
\revision{
We now discuss how \texttt{\TOOLNAME} constructs the execution tree at Alice's end. As per Definition~\ref{def:extree}, an execution tree merges reusable program states of different versions into a single node in the tree. Thus given $n$ audited versions, possibly of different length, to construct an execution tree, we need to identify which states are reusable. Given the per cell  values of state computation time $\delta_i$ and size $sz_i$, state lineage $g_i$, and state code hash $h_i$, we use the following conditions to to identify a reusable state:

\begin{definition}[State equality]
\label{def:equality}
Given two program versions $L_1$ and $L_2$, state $ps_i$ in $L_1$ is equal to state $ps_j$ in $L_2$,  denoted $ps_i = ps_j$, \textit{if and only if}
\begin{itemize}[noitemsep]
    \item  $h_{i} = h_{j}$,
    \item $g_{i} =  g_{j}$,
    \item $\delta$ and $sz$ costs are similar.
\end{itemize}
\end{definition}
In other words we say that two states are reusable if they are equal i.e., they are (i) equal at code syntactic level, (ii) after cell execution, result in the same  state lineage (note state lineage of $i^{th}$ cell depends on state lineage of previous cell), 
and (iii) have roughly similar execution costs.  Program state does not remain equal when cell code is edited, which changes the hash value  of that cell and any subsequent cell.  Similar states across versions also do not remain equal if costs change drastically, \textit{i.e.}, computed on different hardwares (\textit{viz.} GPU vs CPU). Equating state lineage depends on the granularity at which the system events are audited. Since in \toolname, lineage is audited at the level of system calls, there are some pre-processing steps that are necessary to establish equality, such as  accounting for partial orders, abstracting real process identifiers, and accounting for  hardware interrupts. We describe these issues below.

Lineage equality implies that at end of cell $i$ of version $L_1$, $g_i$ is the same as that at end of cell $i$ of version $L_2$. This is true if and only if the \emph{sequence} of system call events (and their parameters)---till $i$ in $L_1$ and $i$ in $L_2$---exactly match.  But if a cell, e.g., forks a child process, which itself issues system calls, then each version's sequence will contain the parent calls and the child process calls interleaved in possibly different orders.  

In Figure \ref{fig:lineage} the parent process forks a child and then issues a `mem' memory call, and the child process itself issues `exec', `open', and `read' calls.  As the figure shows, it is possible that in the sequence for version $L_1$ the `mem' access is before the `read', while for $L_2$ it is after. If we want to correctly determine that the state in $L_1$ is identical to that in $L_2$ at this point, we need to recognize that the sequence of system calls is an arbitrary total order imposed on an underlying partial order.  The partial order for $L_1$ and $L_2$ is identical, while the total order can differ. 

In our implementation, we essentially reconstruct the underlying partial order when we detect asynchronous computation, and match it to identify equality of program states in different versions. This is achieved by separating the events into PID-specific sequences and then comparing corresponding sequences.  The above comparison can only be established when process identifiers are abstracted to their logical values. Memory accesses cannot be abstracted and we just count the number of accesses in a cell. Comparison must also account for external inputs in addition to system events. As Figure~\ref{fig:lineage} shows the hash of external dataset file `new\_fashion' is changed from `b2e1772' to `6789b34. Thus, the two cells cannot be equated even though the order of system call sequence in $E$ is the same.


A related nuance is due to hardware interrupts.  If $P_1$ experiences a hardware interrupt and $P_2$ does not, we make the safe choice: assume the program states are not equal.  (It is easy to make the opposite choice, by simply ignoring hardware interrupts, if so desired.)
}

\section{Experimental Evaluation}
\label{sec:experiments}



\begin{table*}[th]
\parbox{.55\linewidth}{
\tiny
    \centering
    \caption{Six Real-world Applications}
    \label{tab:real}
    \begin{tabular}{|c|c|c|c|c|c|c|}
    \hline 
     {\bf Dataset:}    & {\bf ML1} & {\bf ML2} & {\bf ML3} & {\bf ML4} & {\bf SC1} & {\bf SC2}\\
     \hline
     {\bf Description} & \begin{tabular}[c]{@{}c@{}}
      Neural \\ Networks\\~\cite{ML1a,ML1b}
      \end{tabular} & \begin{tabular}[c]{@{}c@{}} 
     Stock \\ Prediction\\~\cite{ML2a,ML2b}
      \end{tabular} & \begin{tabular}[c]{@{}c@{}} Image \\ Classification\\~\cite{ML3a} \end{tabular} &
      \begin{tabular}[c]{@{}c@{}} Time-Series \\ Forecast~\cite{ML4a} \end{tabular}&
      \begin{tabular}[c]{@{}c@{}} 
       Gas Market \\ Analysis\\~\cite{SC1a}
      \end{tabular} & \begin{tabular}[c]{@{}c@{}} Spatial \\
      Analysis\\ ~\cite{SC2a,SC2b} \end{tabular} \\
     \hline 
      {\bf Changed parameter} &  \multicolumn{4}{c|}{ \begin{tabular}[c]{@{}c@{}} \revision{models}, hyperparameters, test metrics,  datasets, epochs \end{tabular}} &  \multicolumn{2}{c|} {\begin{tabular}[c]{@{}c@{}} datasets and  input parameters \end{tabular}}\\
      \hline
      {\bf Number of versions}  & 25 & 24 & 32 & 36 &12 & 23 \\
      \hline 
      {\bf Version Length}  & 9 - 13 & 9 & 7 - 8 & 17 & 18 & 33 \\
     \hline 
     {\bf Total \revision{(no-cache)} replay cost (s)}  & 33390  & 298  & 2127 & 10696 & 7126 & 10826  \\
     \hline 
       {\bf Cell compute range (s)} & 0.0005 - 1073 & 0.0003 - 8.5  & 0.008 - 50 & 0.01 - 240 & 0.0003 - 926  & 0.0002 - 224  \\
      \hline 
       {\bf Total checkpoint size (GB)} & 57  & 37  & 106 & 566 & 13 & 14 \\
      \hline 
      {\bf Cell checkpoint size (GB)} & 0.2 - 1.8 & 0.2 - 0.38  & 0.4 - 2 & 1.3 - 11 & 0.077 - 0.100  & 0.040 - 0.050 \\
      \hline 
      
    \end{tabular}
}
\hfill
\parbox{.35\linewidth}{
\tiny
    \centering
    \caption{Three Synthetic Datasets}
    \label{tab:synthetic}
    \begin{tabular}{|c|c|c|c|c|c|}
    \hline 
     {\bf Dataset:}    & {\bf CI} & {\bf DI}  & {\bf AN} \\
     \hline
     {\bf Description:}    & {
     \begin{tabular}[c]{@{}c@{}}
     Compute-\\
     intensive
     \end{tabular}} & {\begin{tabular}[c]{@{}c@{}} Data-Intensive\end{tabular}}  & { Analytical} \\
     \hline 
    \textbf{Max. Branch-out Factor}   & 4 & 4 & 4 \\
     \hline
      \textbf{\revision{Max.} Version Length}  & 6 & 6 & 6 \\
      \hline
      \textbf{Number of versions}  & 20 & 20 & 20 \\
      \hline 
    \textbf{Total \revision{(no-cache)} replay cost (s)}   & $\sim$20000 & $\sim$5000 & $\sim$20000 \\
     \hline 
     \textbf{ Cell compute range (s)} & 100 - 600 & 100  & 100 - 600  \\
      \hline 
      \textbf{Total storage size (GB)}  & $\sim$22  & $\sim$18 & $\sim$18  \\
      \hline 
      \textbf{Cell checkpoint size (GB)} & 0.5 & 0.1 - 0.6  & 0.1 - 0.6 \\
      \hline 
    \end{tabular}
}
\end{table*}



{
\begin{figure*}
        \centering
        \begin{tabular}{ccc}
        \begin{subfigure}[b]{0.28\textwidth}
                \centering
                {%
                \includegraphics[width=0.8\textwidth]{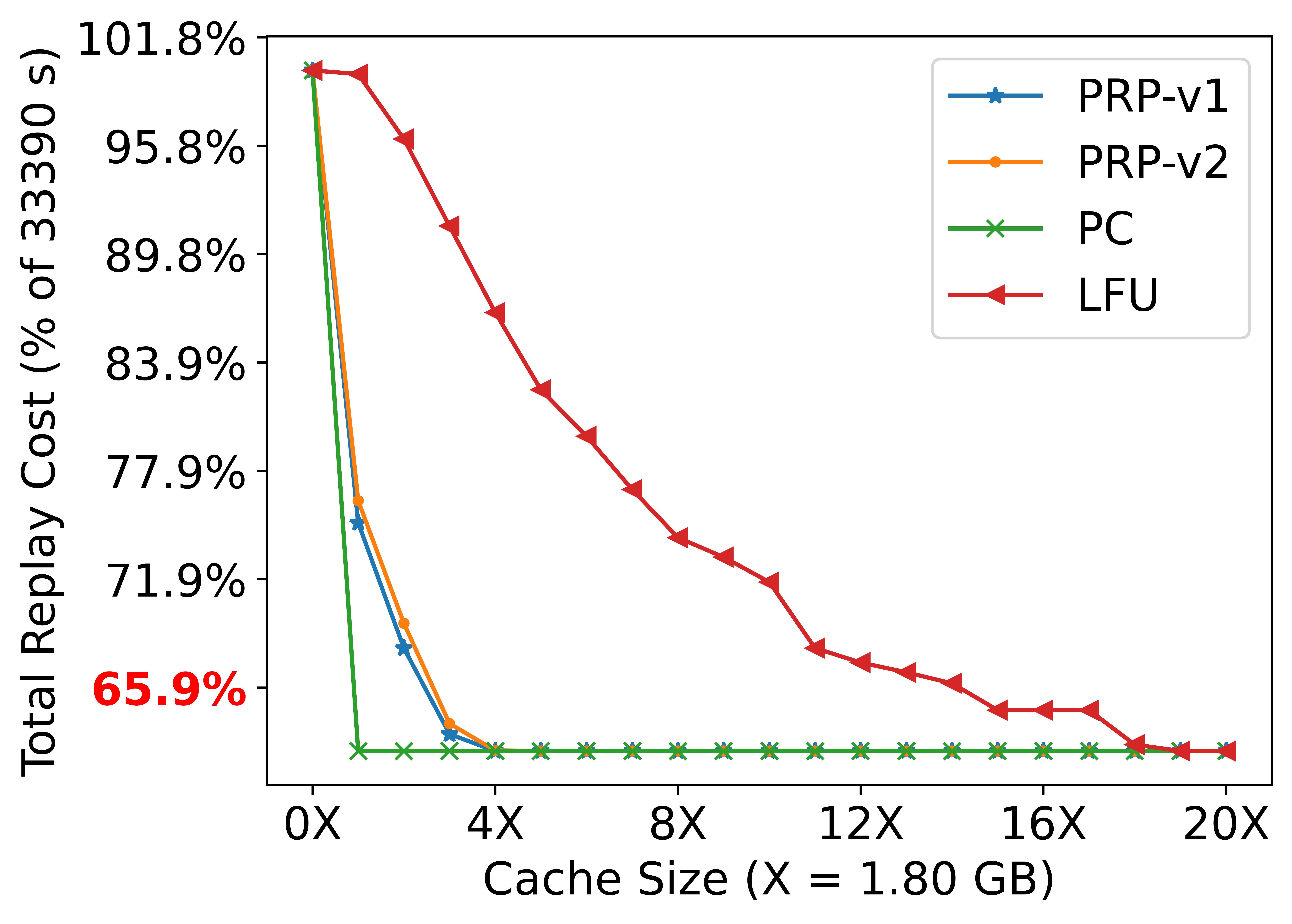}
                }%
                \vspace{-5pt}
                \caption{ML1}
                \label{tempc}
        \end{subfigure}%
        &
        \begin{subfigure}[b]{0.28\textwidth}
                \centering
                {%
                \includegraphics[width=0.8\textwidth]{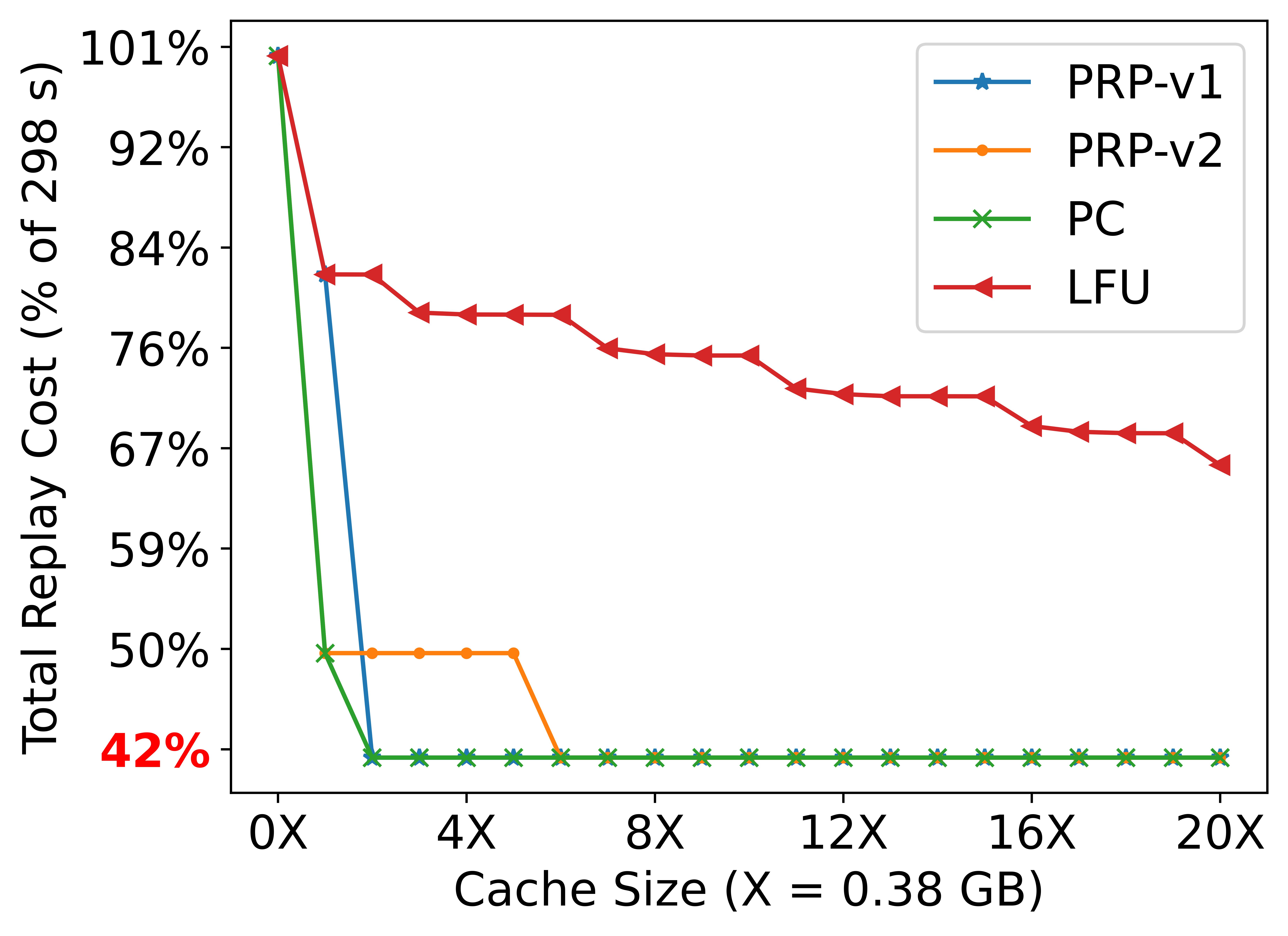}
                }%
                \vspace{-5pt}
                \caption{ML2}
                \label{tempf}
        \end{subfigure}
        &
        \begin{subfigure}[b]{0.28\textwidth}
                \centering
                {%
                \includegraphics[width=0.8\textwidth]{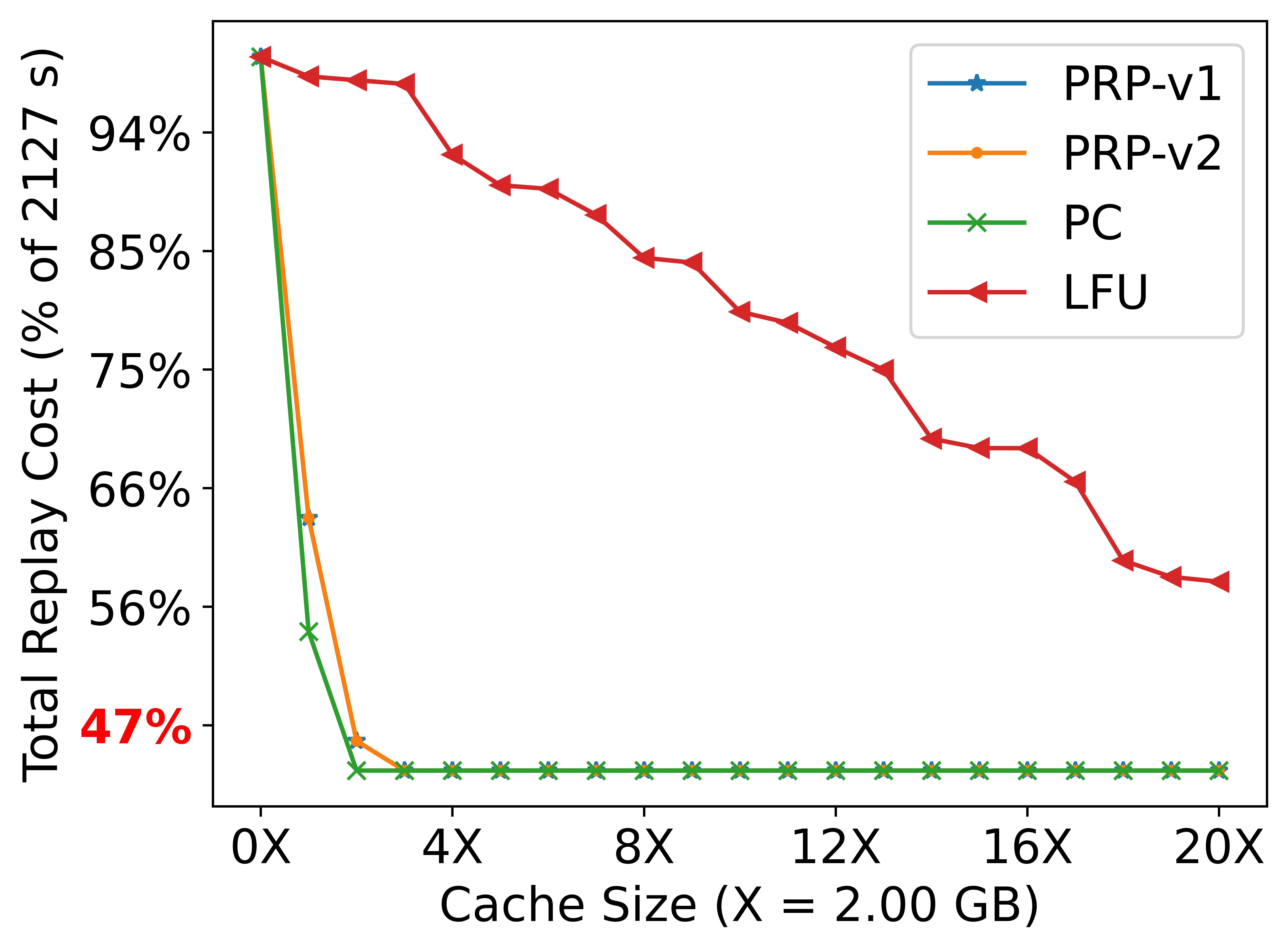}
                }%
                \vspace{-5pt}
                \caption{ML3}
                \label{tempi}
        \end{subfigure}
        \\
        \begin{subfigure}[b]{0.28\textwidth}
                \centering
                {%
                \includegraphics[width=0.8\textwidth]{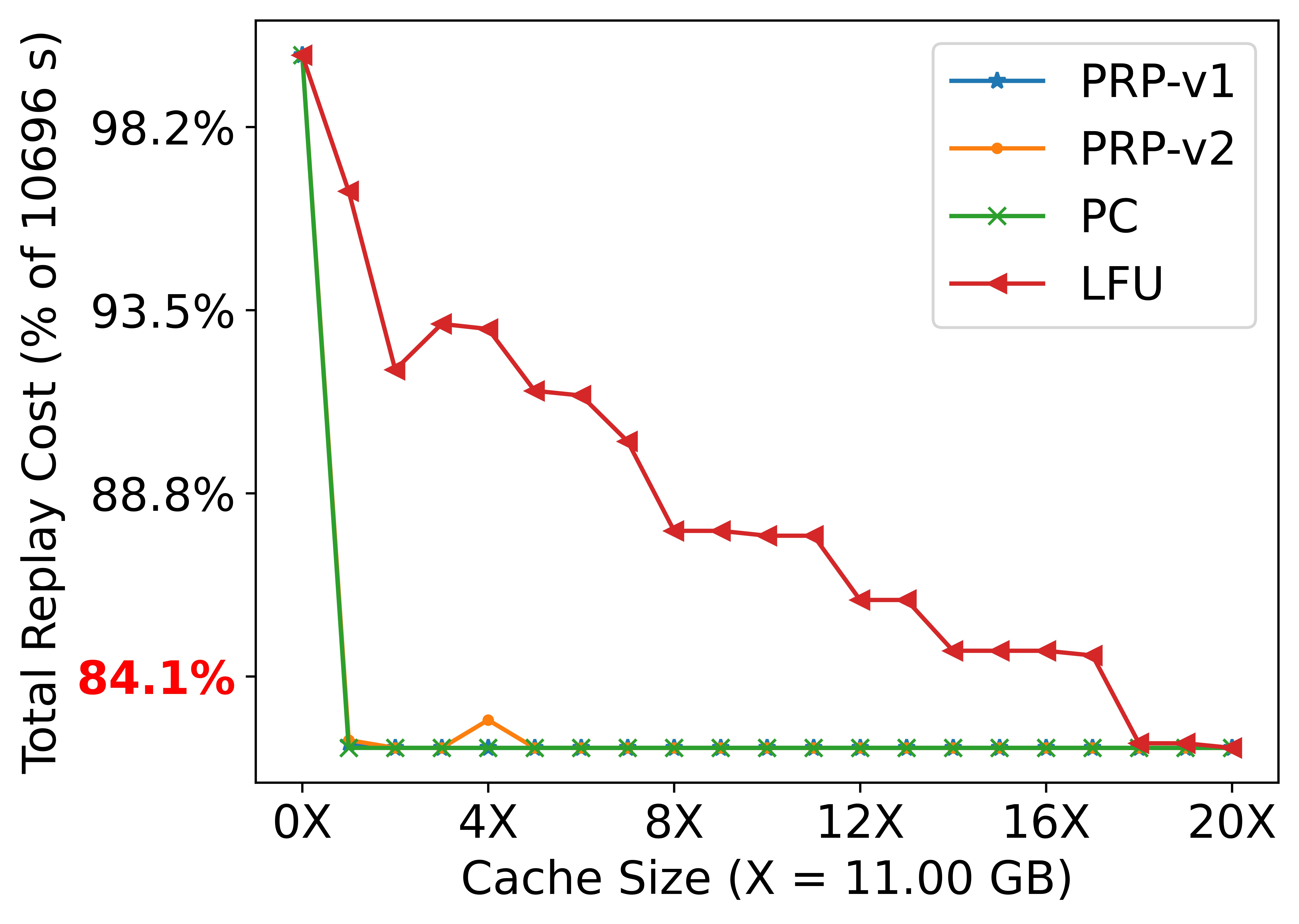}
                }%
                \vspace{-5pt}
                \caption{ML4}
                \label{templ}
        \end{subfigure}
        &
        \begin{subfigure}[b]{0.28\textwidth}
                \centering
                {%
                \includegraphics[width=0.8\textwidth]{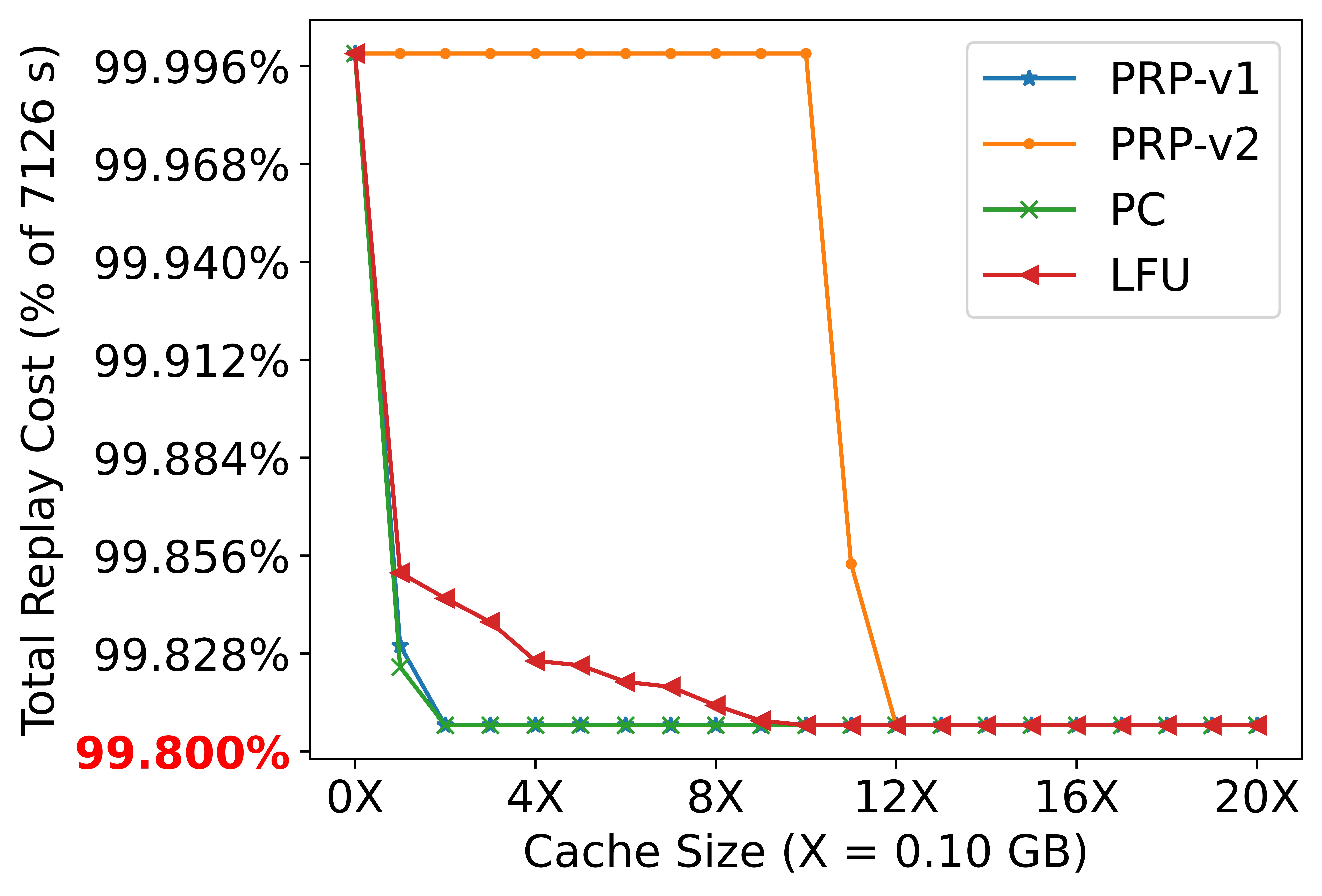}
                }%
                \vspace{-5pt}
                \caption{SC1}
                \label{tempo}
        \end{subfigure}
        &
        \begin{subfigure}[b]{0.28\textwidth}
                \centering
                {%
                \includegraphics[width=0.8\textwidth]{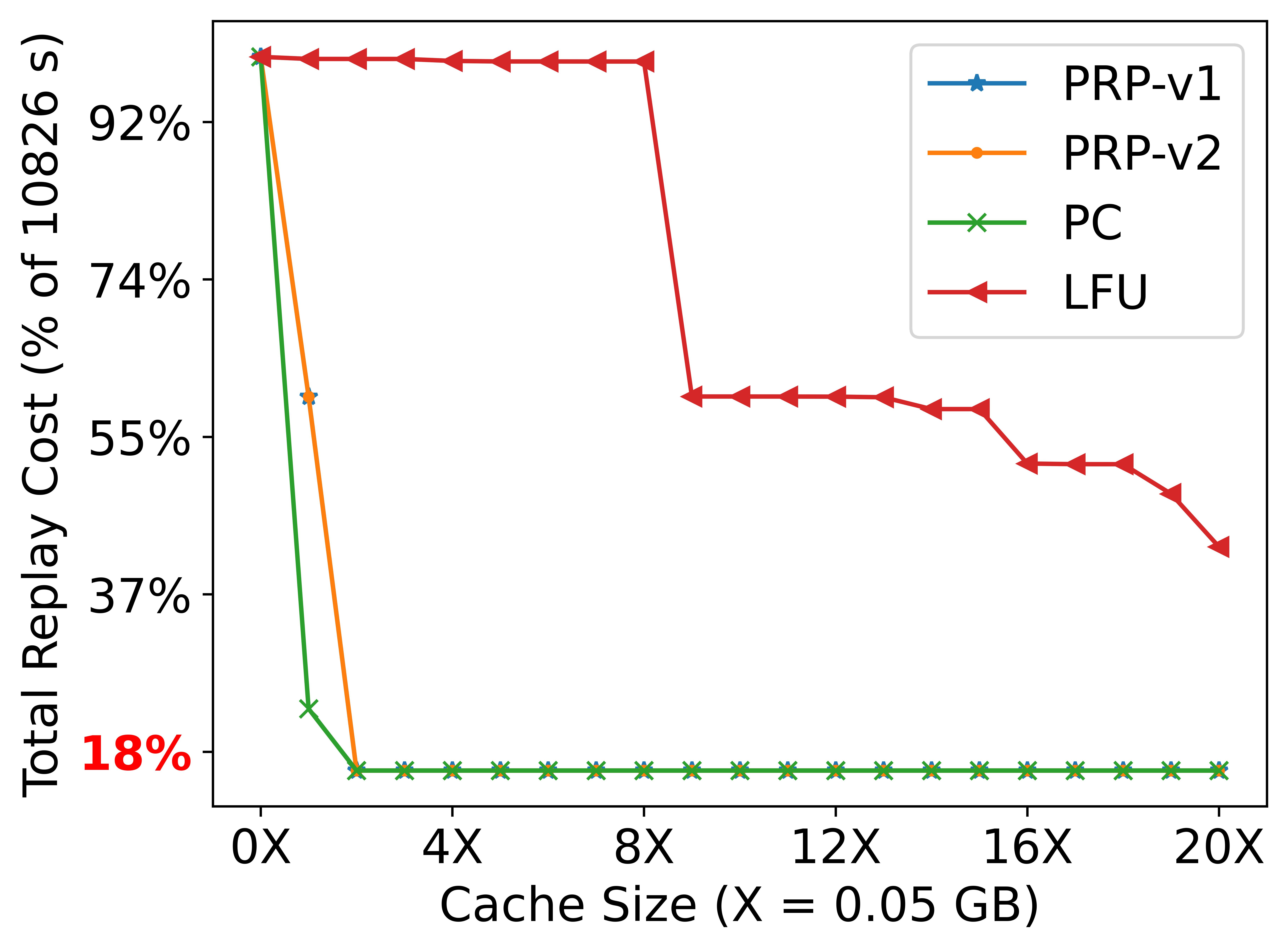}
                }%
                \vspace{-5pt}
                \caption{SC2}
                \label{tempr}
        \end{subfigure}
        \end{tabular}
        \caption{\revision{Performance of DFS algorithms on 6 real-world applications. $X$ denotes the size of the largest checkpoint cell as specified in the last row of Table~\ref{tab:real}. The $y$-axis is truncated to show finer performance variations between algorithms.}}\label{fig:real-perf}
\end{figure*}
}

We now \revision{describe \toolname's implementation and present an extensive evaluation of \texttt{\TOOLNAME} for multiversion replay.}

\vspace{5pt}
\revision{\noindent {\bf Implementation.}
 \texttt{\TOOLNAME} is implemented in C and Python. \toolname relies on Sciunit~\cite{SciunitI:2017:sciunit.run} for monitoring the application on Alice's side and it relies on Checkpoint/Restore in Userspace (CRIU)~\cite{CRIU} to checkpoint/restore program states. \toolname maintains a ramfs cache to maintain checkpoints. These checkpoints are of the process corresponding to the REPL program and not of the container that Sciunit creates. 
 
 We use CRIU as a checkpointing mechanism. This is precisely to enable checkpoint of a process independent of its programming language\footnote{\revision{Native serializations, \textit{viz.} Pickle, provide only a slight performance benefit (1-2\%)}.}. CRIU does not freeze the state of the container but just the application process. Currently, \toolname is integrated with the \revision{I}Python kernel. In future, we plan to integrate  \texttt{\TOOLNAME} with Xeus~\cite{xeus},  which will help us extend  \texttt{\TOOLNAME} to C programs as well. For the purposes of reproducibility we have made available the code for the \revision{audit} and replay mode of \texttt{\TOOLNAME} at ~\cite{chex-code}. We have also provided data sets comprising a set of notebooks with pre-computed execution trees.
}




 We used a combination of real-world applications and synthetic datasets for \revision{evaluation. We ran all our experiments on a 2.2GHz Intel Xeon CPU E5-2430 server with 64GB of memory, running 64-bit Red Hat Enterprise Linux 6.5. The heuristics were developed in Python 3.4.}

\vspace{5pt}
\noindent{\bf Real-world Applications.}
We searched GitHub and identified compute- and data-intensive notebooks, i.e., the programmer had already divided the code into cells. Most of these notebooks were published as artifacts in specific domain conferences (pre-established to be reproducible), and they were described as compute- and data-intensive. 

 We used \revision{four} neural network machine learning applications ({\bf ML}) and \revision{two} scientific computing ({\bf SC}) applications. 
Table~\ref{tab:real} describes the characteristics of these notebooks.
For the majority of the applications, \textit{the number of versions} was determined in consultation with the notebook authors, by identifying meaningful changes to parameter values. Other notebooks were changed similarly. \revision{\textit{Total replay cost}} is the time to run all the versions with no cache. \revision{\textit{Total checkpoint size}} is the space required if each cell of the corresponding execution tree is checkpointed. \revision{\textit{Cell compute range}} and \revision{\textit{Cell checkpoint size}} represents the 
range of cell compute time and checkpoint size ranges\revision{, respectively}. The \textit{changed parameter} row mentions application parameters that were changed to create  versions.
The only way we created versions was by changing parameters. We did not modify any part of the programs in any other way. 

The case of the parameter \textit{epochs} in \textbf{ML} notebooks is special. 
In our case, the \textbf{ML} notebooks embed deep neural networks, in which typically the compute-intensive part is the back propagation during the training phase. Back propagation is usually implemented as an iterative for-loop, whose  upper bound is defined by the \textit{epochs} parameter. Changing {\em epochs} will change the training length and the number of iterations in the for-loop. Such a change to create a new version, however, will also re-run the entire training phase again, which will include the training iterations performed in the previous version. Therefore to create a \revision{ version} when the change is to {\em epochs}, we do not modify the value in-place. Instead we add a new cell. This cell consists of the author-provided training loop but with a incremental range of epochs starting from the last epoch value of the previous cell. This way of modifying the epoch parameter introduces no change to the code and corresponds to incremental training, which is often used in ML to take advantage of previous computations.  

\mysubfigthreebox{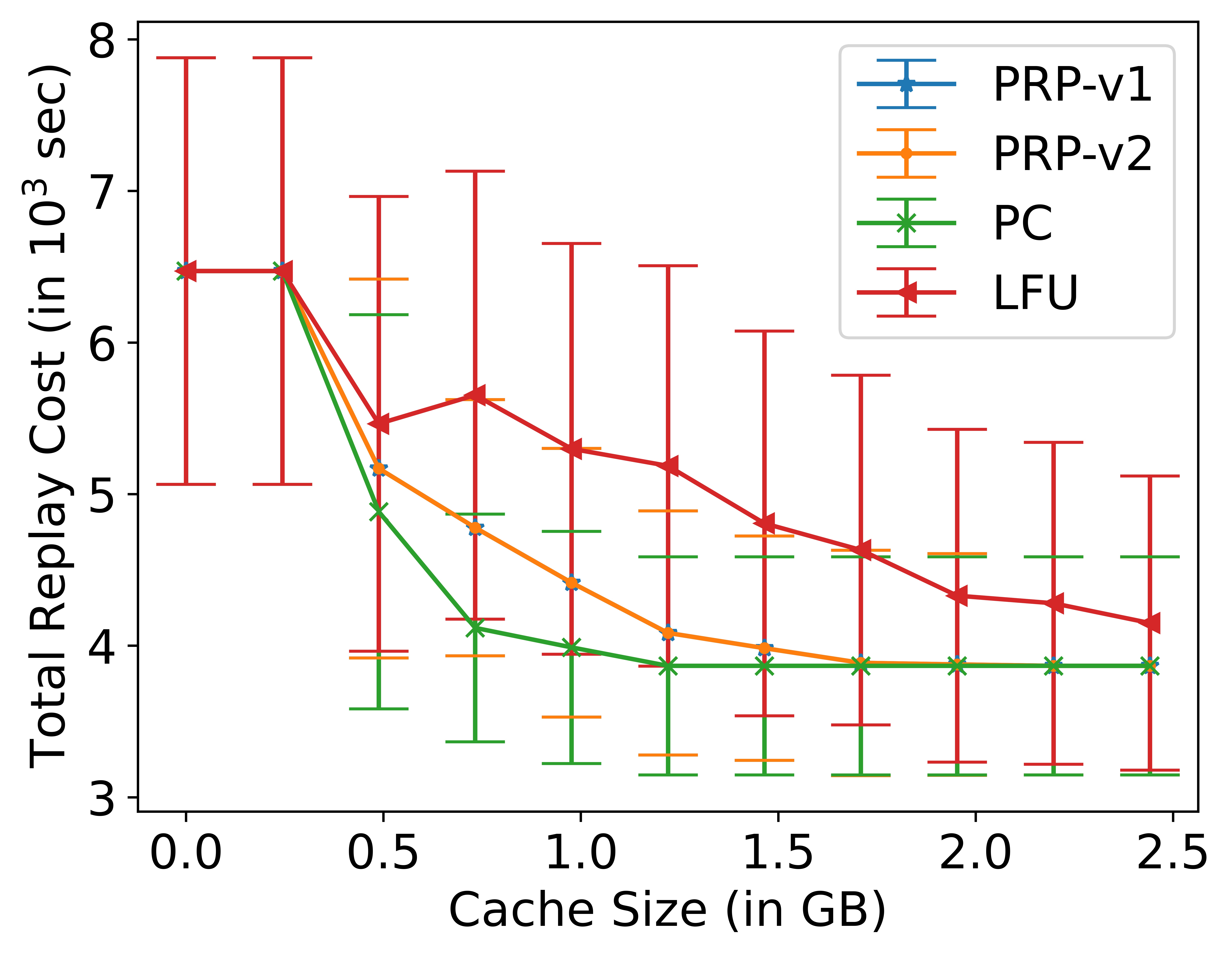}{Compute-intensive (CI)}{fig:ci}{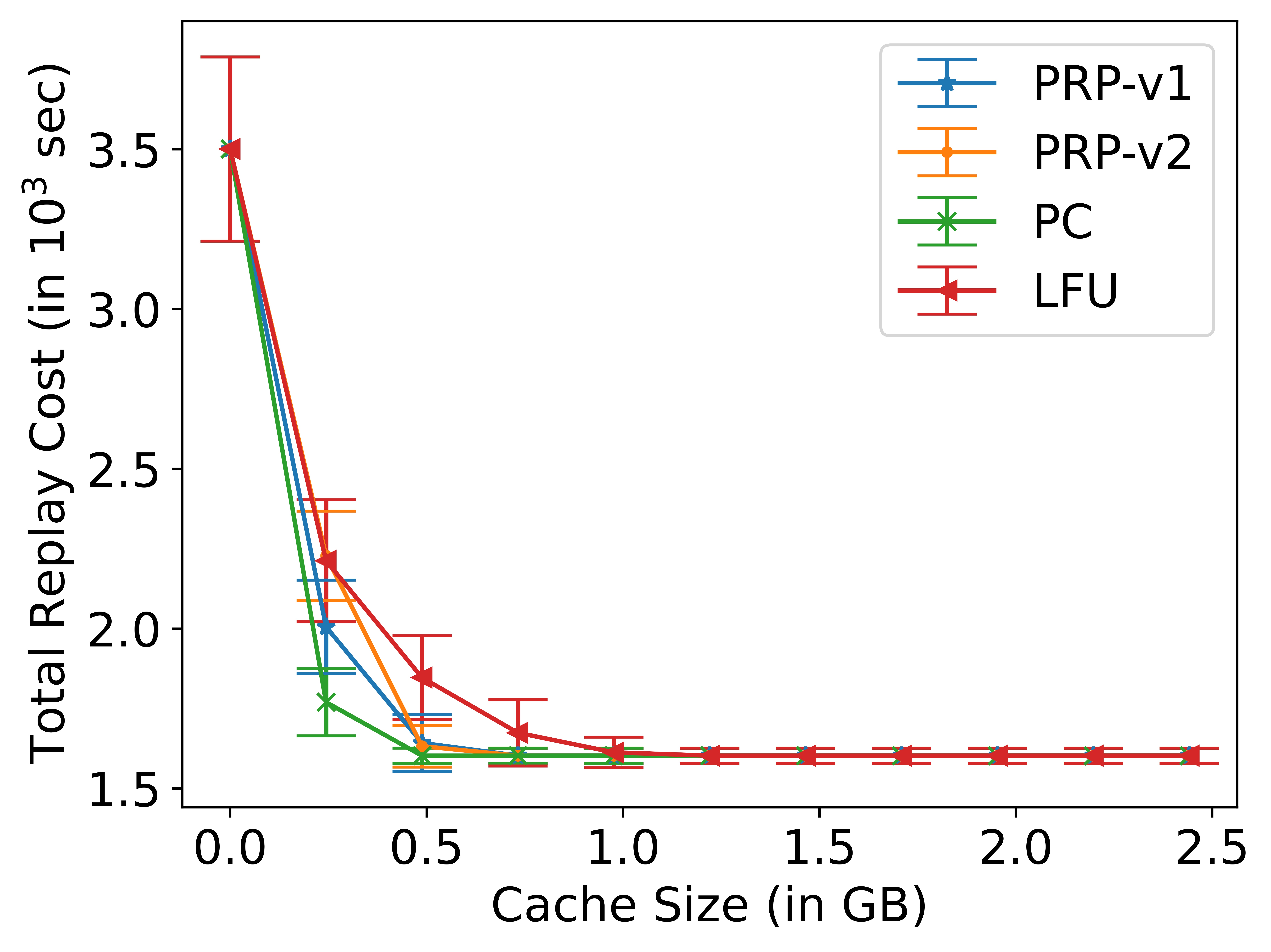}{Data-intensive (DI)}{fig:di}{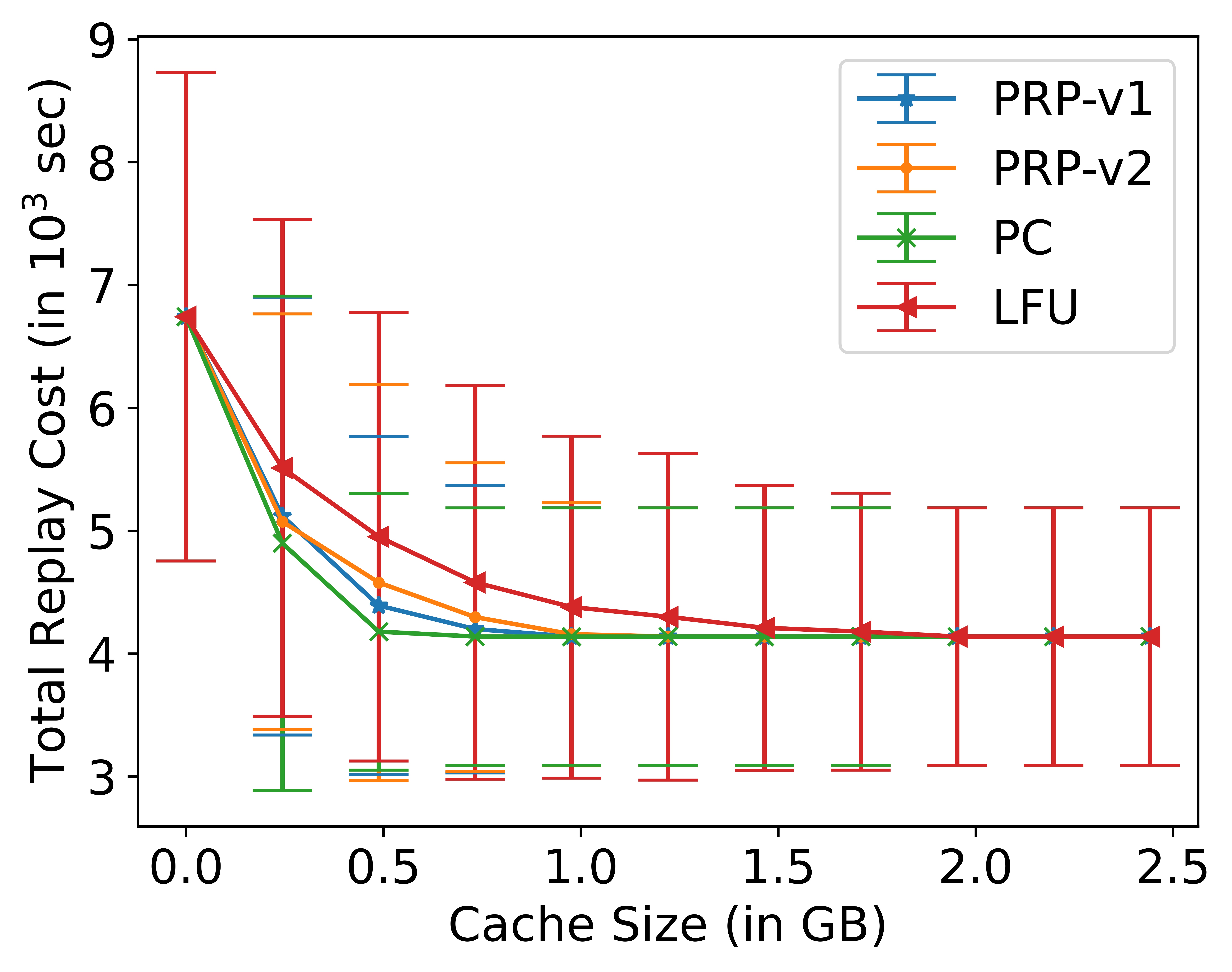}{Analytical (AN)}{fig:an}{Performance of algorithms on 3 synthetic datasets}{fig:synth-perf}


\noindent {\bf Synthetic.} 
To test the sensitivity of our heuristics we randomly generated synthetic execution trees with different costs and sizes. We controlled the tree structure using the following parameters: 

\noindent{\em max. branch out factor}: The maximum number of branches possible at a node. \revision{Each branch is constructed with a 50\% probability.
This leads to trees in which many nodes have a single child. This is what we have observed in real notebooks. }

\noindent{\em{\revision{max.} version length}}: The number of cells in each version. \revision{In general, the length for each version is different because of the randomization described above.} 

\noindent{\em max. number of versions}: The number of leaves in the execution tree \revision{generated by using {\em max. branch out factor} and \em{max. version length}}.
Using the above parameters, we generate three synthetic datasets:
\begin{itemize}[leftmargin=*]
    \item Compute intensive  (\textbf{CI}): In the \textbf{CI} tree, the compute  cost ($\delta$) of cells is high and the checkpoint cost ($sz$) is modest.
    \item Data intensive (\textbf{DI}): In the \textbf{DI} tree, the checkpoint cost ($sz$) of cells is high and compute cost ($\delta$) is modest.
    \item Analytic  (\textbf{AN}): In the \textbf{AN} tree, compute and checkpoint costs \textit{i.e.,}  $\delta$ and $sz$ increase with \textit{version length}.
\end{itemize}
Table~\ref{tab:synthetic} presents the total compute time and total storage size as well as the compute and storage ranges per cell. 

\vspace{5pt}
\noindent {\bf Baselines.} \revision{ IncPy~\cite{Guo:TaPP:2010:IncPy,Guo:ISSTA:2011:IncPy} avoids recomputing a function with the same inputs when it is called repeatedly or across program versions.} Despite our best attempts we could not get IncPy to run with our real datasets. IncPy is not longer actively maintained and is Python 2.7 based which creates conflicts with more recent notebooks. We simulated the Vizier system, by taking one notebook version at a time~\cite{brachmann2020your},and using the simple caching policy that is used for Vizier: Least Frequently used (\textbf{LFU}), which is a standard caching algorithm. We adapt \textbf{LFU} to our case by checkpointing every cell of the first version of a notebook till the cache space fills up.
As subsequent versions arrive, the cache  eviction policy is decided by the measure $frequency \times \#\ of\ nodes\ in\ subtree/cell\ size$, i.e., retaining cells which are used frequently and are responsible for larger subtree, normalized by their size. Least recently used, another standard caching algorithm, is not relevant in our case due to the depth-first replay order.

\subsection{Experiments}
\revision{We first evaluate the benefit different algorithms provide in terms of reduction in replay time. We then evaluate the overhead of operating \toolname.}

\subsubsection{{\revision{\bf Comparing decrease in replay cost via different algorithms}}}
\label{sec:replaycost}
\revision{Persistent Root Policy (\textbf{PRP}) and Parent Choice (\textbf{PC}) make different choices with respect to cells that must be retained in cache for recomputation. In this  experiment, we evaluate how those decisions compare with the (\textbf{LFU}) baseline.}\delete{We operate different sized caches as per the Persistent Root Policy (\textbf{PRP}) and Parent Choice (\textbf{PC}) heuristics, and compare them on real-world applications and synthetic datasets.} Recall that {\bf PRP} has two versions: \revision{\textbf{PRP-v1}, in which we cache checkpoints  greedily based on contribution to reduction in cost, and \textbf{PRP-v2}, in which we normalize the cost reduction by the ccheckpoint size.}


\revision{To compare algorithmic performance, we choose a cache size that is equal to the largest checkpoint size in a notebook and compute total replay time. The $y$-axis is initialized with a non-zero value to show finer comparisons between algorithms. For both \textbf{PC} and \textbf{PRP} algorithms on real-world applications, as is expected, Figures~\ref{fig:real-perf}(a)-(f), show decreasing compute times ($y$-axis) as the cache size is increased ($x$-axis).}\delete{For all real-world applications,
Figures~\ref{fig:real-perf}(a)-(d) and (f), show decreasing compute times (y-axis) as the cache size is increased (x-axis) when cache is operated as per all the algorithms.} We \revision{also} see  \textbf{PRP} and \textbf{PC} always perform substantially better than \textbf{LFU}, and \textbf{PC} reduces total compute cost more than either of the \textbf{PRP} versions. 

\revision{Both these result trends are not exhibited in Figure~\ref{fig:real-perf}(e) \revision{(\textbf{SC1}) and, to an extent, in Figure~\ref{fig:real-perf}(d) \textbf{ML4}}. In (e), as we observe, none of the algorithms, including the baseline {\bf LFU}, show any  benefit of caching. This is because in this notebook only the last cell of each version is compute-intensive, and none of the intermediate cells are cache-worthy. In (d), similarly, most computation is towards the later cells; \textbf{PRP} and \textbf{PC} still find some ways to optimize which \textbf{LFU} cannot find. The effect of reuse of intermediate results is well-demonstrated when comparing {\bf ML4} and {\bf SC2} which exhibit similar total replay costs. However, there is a much greater reduction in total replay cost in {\bf SC2} (from 100\% to 18\%) as there are several compute-intensive pre-processing steps in the earlier cells of the notebook, where as in {\bf ML4} most computation occurs towards the later cells. 

Analyzing deeper we also observe these trends: (i) Sometimes, initially, \textbf{PRP} performs better, and this happens due to small cache size effect, since \textbf{PC} becomes a clear win with some additional cache space;} (ii) \textbf{PRP-v1} performs better than \textbf{PRP-v2} indicating that eviction on a cost/size measure leads to more greedy eviction policy where checkpoints are evicted which need to recomputed later; and finally (iii) \textbf{ML1}, \textbf{ML3} and \textbf{SC2} are compute-intensive notebooks. \revision{Using the \textbf{PC} algorithm, these notebooks show a reduction of 60-65\% in their compute time at a size of the cache which is at most double the size of the largest checkpoint cell in the notebook. This indicates that smart algorithms can provide significant benefits even with small cache sizes.}  
\delete{We also observe the following: (i) \textbf{PC} reduces total compute cost more than either of the \textbf{PRP} versions, except initially in Figure~\ref{fig:real-perf}(b) and(d). We attribute the initial high cost owing to small cache size effect, since \textbf{PC} becomes a clear win with some additional cache space; (ii) \textbf{PRP-v1} performs better than \textbf{PRP-v2} indicating that eviction on a cost/size measure leads to more greedy eviction policy where checkpoints are evicted which need to recomputed later; and finally (iv) \textbf{ML1}, \textbf{ML3} and \textbf{SC2} are compute-intensive notebooks. Using the \textbf{PC} algorithm, these notebooks show a reduction of 60-65\% in their compute time at a size of the cache which is slightly less than the average-size of their respective notebooks 
This indicates the benefits of optimizing on cache space with smarter replay orders.} 
\delete{We obtain similar results for synthetic datasets, and, for lack of space, only include the figures for synthetic results in the extended version of the paper~\cite{CHEX-techreport}.}
\revision{Figures~\ref{fig:synth-perf}(a)-(c) show similar result for synthetic datasets. We note from Table~\ref{tab:synthetic} that 0.6GB is the minimum amount of cache needed to cache any cell. Thus the behavior of algorithms before 0.6GB does not reflect optimal decisions but we include it as cells of size less than 0.6GB may still be cached.  Since 
\texttt{\TOOLNAME} caches checkpoints in memory, results show  
an advantage of \textbf{PC} over \textbf{PRP}. The advantage continues for  cache sizes which are smaller in size than an average sized notebook. The algorithms advantage wanes off for larger sized caches due to all relevant cells cached owing to the structure of the simulated execution tree.} 

\subsubsection{\revision{\bf Determining number of versions replayed with fixed cache size}} 
\begin{figure}[h]
    \centering

    \includegraphics[width=0.45\linewidth]{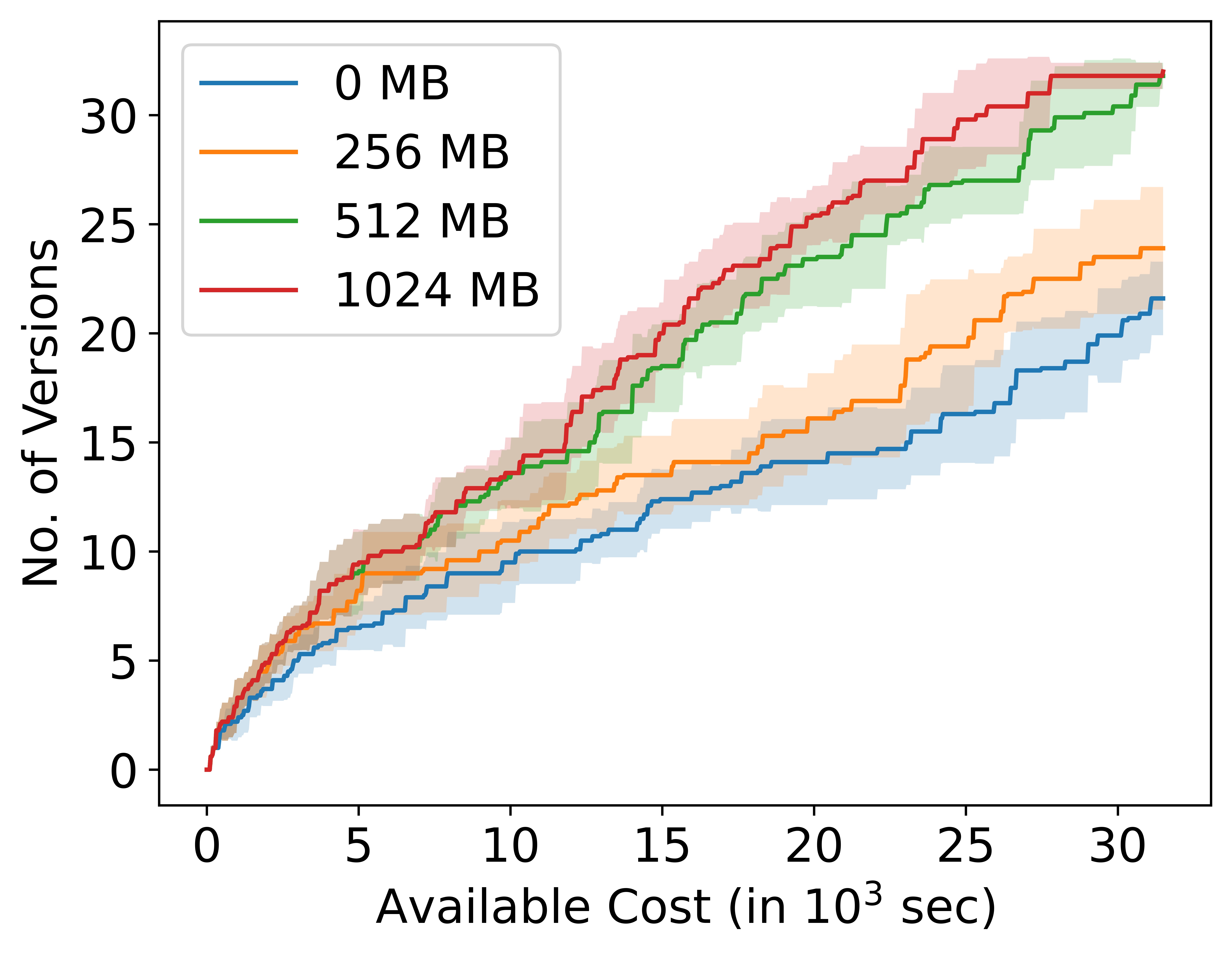}
        \vspace{-5pt}
    \caption{For given cache sizes, number of versions that can be replayed with \toolname in a given amount of time, on the {\bf AN} dataset.}
    \label{fig:versions}
    \vspace{-5pt}
\end{figure}

We also examine the direct benefit of a system like \texttt{\TOOLNAME} for users. For most users \texttt{\TOOLNAME} will be configured with a given amount of \revision{cache} space. Users, however, have time constraints. Thus we determine, for given cache sizes, number of versions that can be replayed with \toolname in a given amount of time, on the {\bf AN} dataset. 
 Figure~\ref{fig:versions} presents the result (number of versions ($y$-axis) for the amount of time it takes to replay them ($x$-axis)) for a \revision{given cache size, the value being either: no cache, 0.25GB, 0.5GB, and 1GB}. The Figure shows that a user can run 50\% more number of versions  by doubling the space for the same fixed amount of time. To be able to run larger number of runs for the same amount of time has implications for scaleable collaborative sharing and artifact evaluation use cases. 

\vspace{-10pt}

\begin{figure}[h]
    \centering

\includegraphics[width=0.45\linewidth]{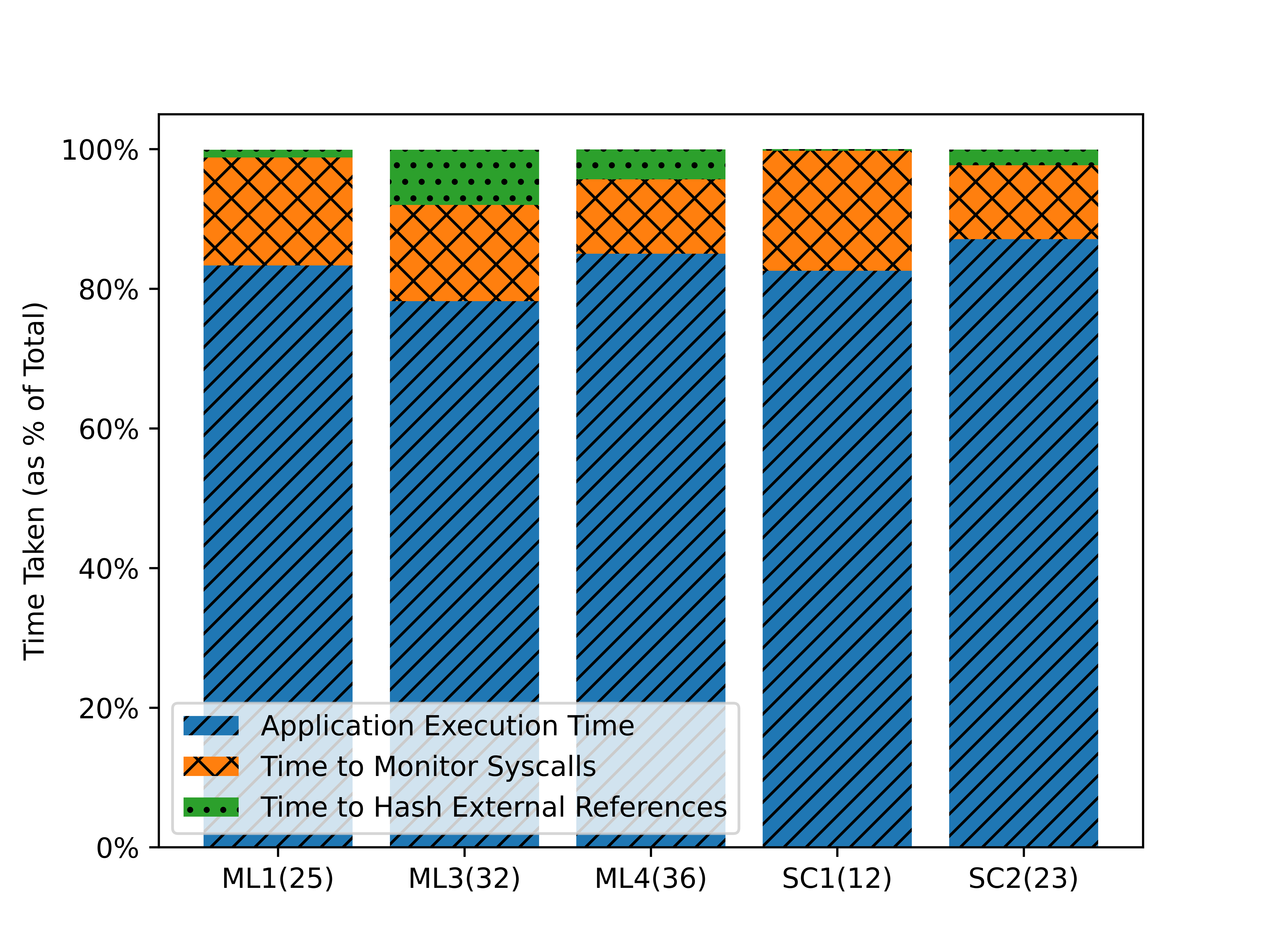}
        \vspace{-5pt}
    \caption{\revision{The overhead of auditing $\delta, sz,g$ and $h$ in real world applications with $>$ 5 minutes of replay cost. }}
    \label{fig:auditing}
    \vspace{-5pt}
\end{figure}


\mysubfigthreebox{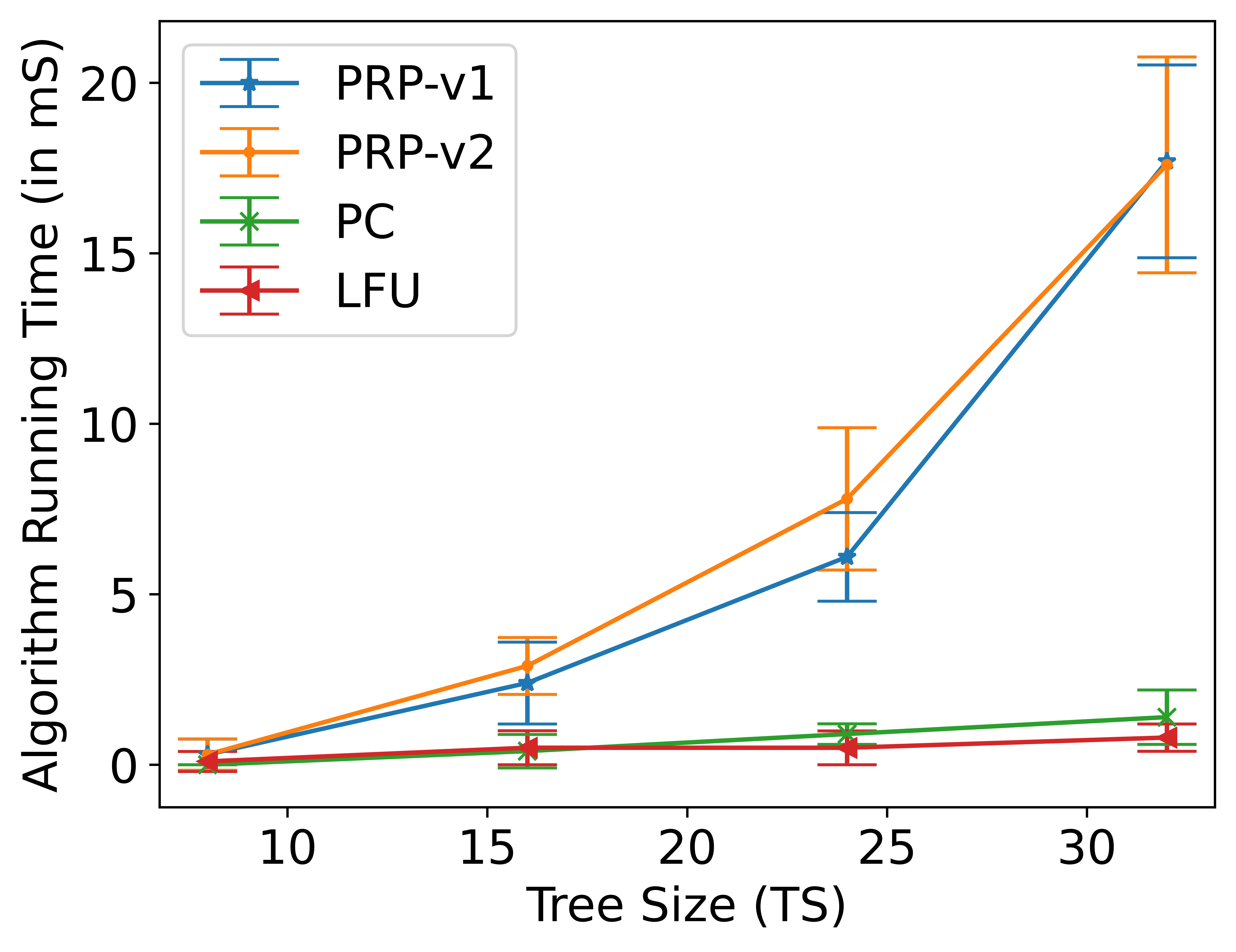}{}{fig:run}{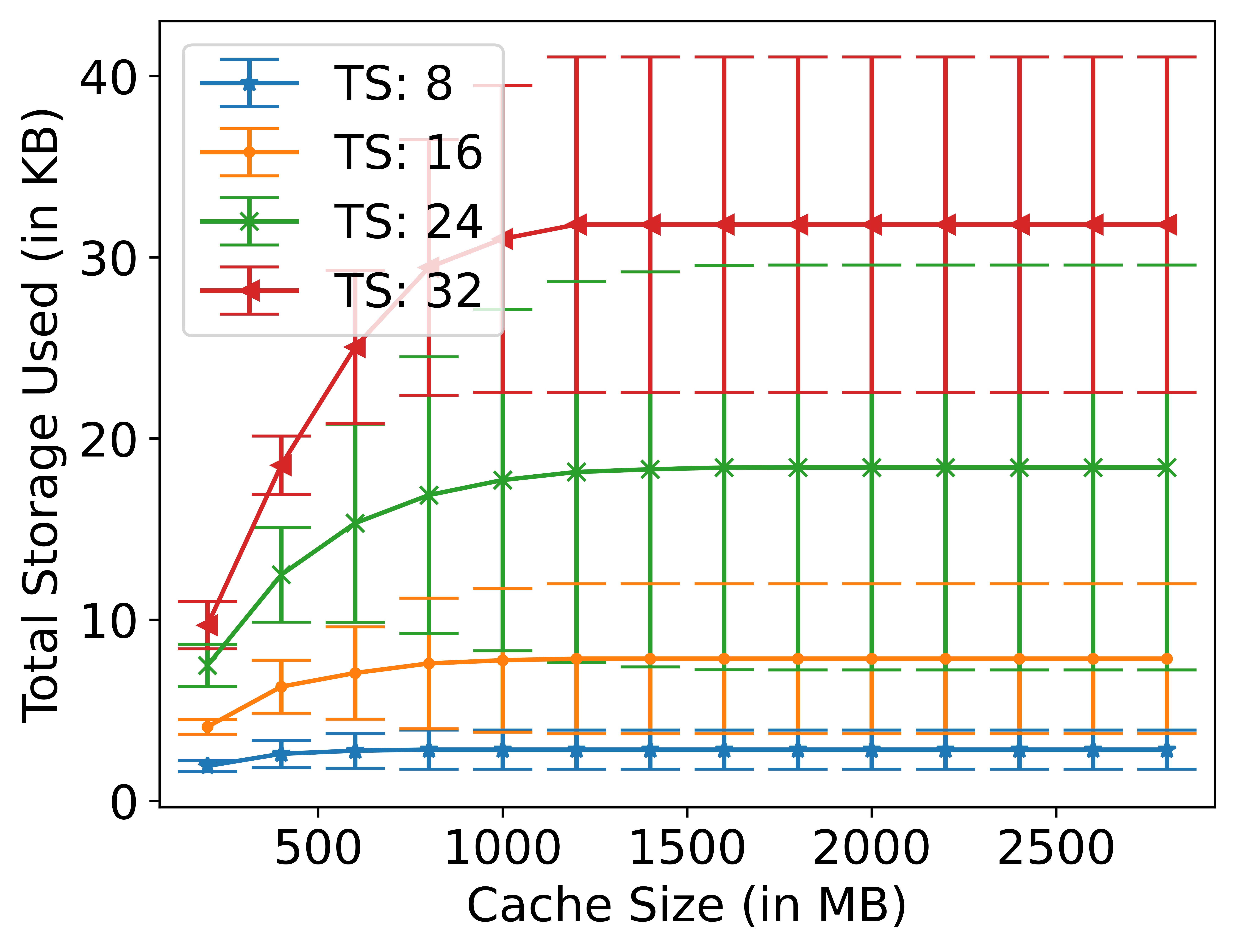}{}{fig:storage}{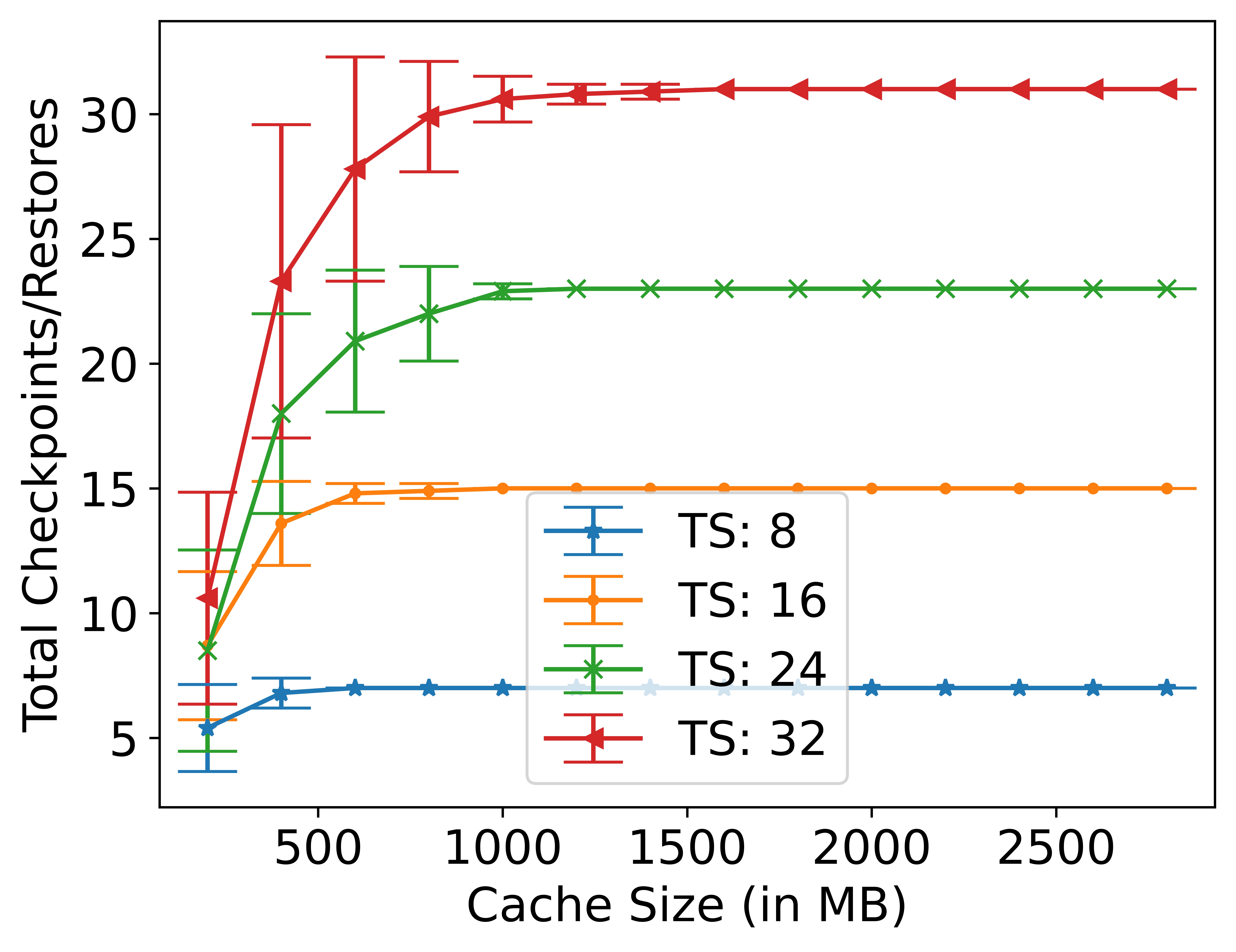}{}{fig:cr}{Algorithm complexity \revision{of \textbf{AN} workloads}: (a) running time, (b) storage size \revision{for {\bf PC}}, and (c) number of checkpoints/restore-switch \revision{for {\bf PC}}}{fig:complexity}

\subsubsection{\revision{Time and space required to run \toolname}}
\label{sec:timeandspace}
\revision{
We first determine the  cost of auditing an application in \toolname.

 \vspace{5pt}
\noindent{\bf Cost of Auditing.}
\toolname performs auditing of state for 
each version of an application in terms of  computation time $\delta$, state size $sz$, state code hash $h$, and state lineage $g$. We report both normal execution and audited execution as a percentage of the total time of using \toolname on a real application.

Amongst these audited quantities, the primary overhead is the additional time required to audit the application for state lineage, i.e., $g$. We further divide  time to audit for $g$ into time required to (i) monitor and log system events in the application, and (ii) the time required to compute the hash of any external content that is referenced. 
As Figure~\ref{fig:auditing}(a) shows, a 15-25\% of total  auditing overhead is added across all applications. We are reporting 5 out of six applications as \textbf{ML2} has relatively insignificant running time to begin with.

Also the time to perform cell equality and construct the execution tree is negligible.
Alice shares a package with the execution tree, the size of which is less than 1KB.
}

\vspace{5pt}
\revision{\noindent{\bf Cost of computing cache eviction decisions.}
We have shown the multiversion replay problem to be NP-hard; \textbf{PRP} and \textbf{PC} are heuristic algorithms, and thus have some time and space cost of making the cache eviction decisions.}
In \revision{this experiment}, we measure the \revision{cost}\delete{overhead} of \revision{using} \textbf{PRP} and \textbf{PC} algorithms \revision{in comparison to LFU} in terms of running time, space \revision{used}, and \revision{number of times}\delete{overhead of} checkpoint/restore \revision{call was made}. \revision{We experimented with the \textbf{AN} synthetic dataset.}

\revision{The variability in running time of the algorithms with cache size is negligible (~[0.05\%])}. \revision{Therefore, we fix the cache size to 1GB and show the variation with respect to two other parameters, the number of nodes in the tree, i.e., tree size and the different algorithms.} 
\delete{
We performed the run-time overhead experiment with a 1GB of cache size.} Figure~\ref{fig:complexity} (a) shows that \textbf{PC} is better than \textbf{PRP} in terms on run-time overhead, but in terms of space, it has a cost as shown next.  

\textbf{PRP} has negligible  state maintenance as it uses the execution tree to determine the order. However, 
\textbf{PC} takes storage, because it has to store all possible combinations of execution orders for different cache eviction sizes to get the most optimal one.
Figure~\ref{fig:complexity} (b) shows the increase in storage for different tree sizes as cache size is increased. \revision{Despite these differences, w}e highlight that the runtime and memory overheads of both algorithms is much lower (0.5-2\%) than the overall compute time and storage of multiversion execution of any given real dataset.



The above experiment  measures decision-making time \revision{and space}. In practice to implement the decisions we must account for in-memory checkpoints and restore (C/R) time. In general, time to C/R are proportional to the size of the checkpointed state and are negligible. So we measured the number of times C/R were performed to check if small C/R costs adds to the overall latency of multiversion execution (Figure~\ref{fig:complexity} (c)). As we see C/R costs are negligible, and algorithm decision making accounts for the primary cost of \toolname. 

Apart from the experiments reported above, we attempted a comparison between \textbf{PC} and an optimal algorithm,\revision{using the {\bf AN} dataset for comparison.} For optimal, we wrote our problem, the \probname, as an Integer Linear Program (ILP) and attempted to solve it with the Couenne optimizer~\cite{coin-or}. The Couenne timeout was set as ten minutes. For tree size of 2-6 nodes, Couenne finished finding a solution in less than 10 seconds, but after that the time starts increasing exponentially\revision{.}\delete{ as shown in Figure~\ref{fig:optimal}} \revision{At 12 versions and an execution tree of 20 nodes, the optimal solution could not be found within the set time out. On increasing the number of versions, it took more time to find the optimal solution than naive replay (without cache).} On the other hand, as we show in Figure~\ref{fig:complexity}(a) we took milliseconds to \revision{find a solution} for more than a tree-size of 30.  

\revision{Since we only found optimal solution for small trees, in terms of the quality of the solution, we found the replay cost of {\bf PC} similar to optimal.}\delete{As regards replay cost, we found the replay cost of {\bf PC} similar to optimal for small tree sizes.} For larger tree sizes, we anticipate it to give better cost estimates, but given the large running time of optimal for larger and complex instances, we assess, it is not worth it.  Finally, the overhead of implementing the decisions in \toolname is too small to be measured and often smaller than the variance between multiple runs.




\delete{

}

\vspace{-10pt}
\section{Related Work}
\label{sec:rw}

\delete{This section presents related work in different directions.}

\vspace{4pt}
\noindent {\bf Tools and hubs for sharing and reuse.} \delete{Sharing and replaying is essential for verifying, and reproducing complex applications.} Several user-space virtualization based tools have recently been proposed to enable sharing and repeating computations~\cite{Guo:2011:CDEFull,Janin:CARE,Pham:ICDE:LDV,Chirigati:2016:ReproZip,Ton:2017:Sciunit,o2017engineering}. \revision{These tools do not address multiversion replay.}\delete{
However, when multiple versions of a program are packaged and shared these systems  do not attempt to replay them efficiently.} In a virtualization package, code and data remain separate as files or databases~\cite{Pham:ICDE:LDV}. Computational notebooks, which combine code and data, have received wide attention recently for sharing and use~\cite{Guo2020design}. Notebook sharing, like package sharing, is easy but \textit{(re-)}execution across versions remains sequential. Nodebooks~\cite{Nodebook} and Vizier~\cite{brachmann2020your} are specialized notebook clients that support and store notebook versions at a cell level. Neither, however, compute deltas between versions or trade computation for storage. \delete{In fact we undertook the current work after a conversation with the authors of~\cite{brachmann2020your} in which it became clear that there was an efficiency gap when it came to multiversion programs. In this paper, we have tried to plug that gap by showing how REPL systems can efficiently replay versions by employing data caching.}  Our work complements specialized notebook systems used for interactive development~\cite{koop2017dataflow}, and given lineage from these systems~\cite{macke2020fine}, replay can be enabled.

\noindent {\bf Execution lineage.} There are several provenance models for capturing execution lineage~\cite{Stamatogiannakis-IPAW}. In this paper, we adopt the system-event trace analysis process that is also used in other whole system provenance tracking methods~\cite{gehani2012spade,balakrishnan2013opus,stamatogiannakis2014looking}\revision{.}\delete{, but we do not incorporate more detailed  static analysis and taint flow analysis which further increase time for creating enhanced specification, and often require compile-time instrumentation.} 
 
\noindent {\bf Data caching.} Data management systems have a rich history of employing object caches that tradeoff space for time to improve performance of applications. Semantic caching allows caching of query results~\cite{dar1996semantic,roy2000efficient}, web-object caching allows caching of web objects~\cite{cao1997cost,jin2000popularity}, and query-based object caching allows database object caching based on queries~\cite{malik2005bypass}. In all of these works, the workload sequence is not known. In the multi-query scenarios~\cite{roy2000efficient} the workload is presented as set of queries and hence there is the possibility of caching the results of common sub-expressions and reusing them across queries. However, efficient reuse in the multi-query setting primarily involves searching through the space of query answering plans to identify plans that could potentially lead to optimal reuse. In certain cases not finding the optimal plan and blindly reusing common subexpressions may blow up the computation time because a large join may be required. Our scenario appears similar but we do not have the wiggle room provided by the semantics of a query, nor the potential pitfalls associated with blind reuse\revision{.}\delete{: we simply have multiple versions of a single program. Consequently, the tradeoffs are very different.} 

\noindent {\bf State management for recomputation.} \delete{State management and reuse is vital when recomputing.} ~\cite{to2018survey} provides an excellent survey of state management for \delete{incremental} computation. State can be recomputed from lineage or state can be stored `as-is’. \revision{In SciInc~\cite{Youngdahl19} state is recomputed from lineage that is versioned. Versioned lineage or causality-based versioning~\cite{muniswamy2009causality,Youngdahl19} leads to correct computation of state  for incremental replay. In this work, on the contrary, we are concerned with state that is stored `as-is'. Several works store `as-is' state---this state is state of a variable, query, program, or configuration~\cite{to2018survey}. Similar to ~\cite{kwon2008fault,hwang2007cooperative,hakkarinen2012multilevel,nicolae2013ai}, in this work, our operator is program state. However, in these works the purpose is fault-tolerance, and so the system periodically checkpoints but does not consider space limitations.} We determine  a limited number of checkpoints of program state to save in-memory space, and using lineage, choose to simply recompute when efficient. 
To reduce space an alternative would be to incrementally checkpoint as explored in differential flows~\cite{mcsherry2013differential,murray2013naiad} and query re-opt\-\-\-imization~\cite{liu2016enabling}. These approaches are not extendable to checkpoints of program state, which is an in-memory map. \revision{Very recently checkpointing was used to improve efficiency, but the checkpoint frequency is periodic~\cite{garcia2020hindsight}.
}
\noindent {\bf Checkpoint location.} Deciding when to checkpoint has received attention in HPC  scheduling~\cite{bouguerra2012complexity,robert2012complexity}. \revision{A}\delete{but the} primary objective \delete{there} is to minimize the amount of computation that needs to be redone in case the system \revision{fails}\delete{goes down}. \delete{Besides,} In HPC workflows the checkpoint also has an overhead. We consider machine learning and scientific computing programs in which the checkpoint overhead is nearly zero\revision{.}\delete{, and our setting involves a complete replay of multiversion programs, which makes it quite different.}

Closer in spirit to our work is the DataHubs~\cite{bhattacherjee2015principles} system that seeks to maintain multiple versions of large data sets without fully replicating them. In this system
some versions are stored fully materialized and others are stored only as deltas linked to other versions. The\delete{ir} problem \revision{is to} trade\delete{s} off total storage required versus time taken to recreate a \delete{given} version. At a glance, it is possible to think that the program states of the cells of our multiversion program can be aligned with the data sets \revision{considered in }DataHubs\delete{ considers}. However, the fundamental difference is that DataHubs assumes each version of a data set \delete{(corresponding to a program state in our case)} has {\em already been created} the first time\delete{, and their job is just to store the collection of versions efficiently for recreation}. Thus, they assume that at least one version of the data set is stored in its entirety. In \toolname, the equivalent thing would be for Alice to share some of the program states generated in her execution with Bob. This defeats the entire purpose of independent repetition by Bob.
\vspace{-5pt}
\revision{\section{Discussion}
\label{sec:Discussion}

We now discuss any assumptions that \toolname makes and our results. We assumed that \texttt{\TOOLNAME} works with REPL cells, but, in general, we do not constrain users like Alice to program with REPL interfaces. If the code is not developed via a REPL interface, \texttt{\TOOLNAME} preprocesses it into cells, akin to a program developed via a REPL interface, before monitoring. This preprocessing takes care to not split functions or control flows into separate cells. Thus every input program is automatically transformed into an equivalent REPL program and then entered into the \texttt{\TOOLNAME}. 

We have assumed multiple versions for a given program. We make no assumptions on the types of edits that constitutes a version on Alice’s side. Thus, Alice can change values of parameters, specifications of datasets, models, or learning algorithms. She can also add or delete entire cells. In practice we have found such versions to not correspond to development versions but as separate branches in version-control repositories. In workflow systems they also correspond to independent, but related, experiments. 

We have only demonstrated a scenario in which Alice shares notebooks with Bob for multiversion replay. \delete{While the same situation can be mirrored thus making it possible for Bob to share notebooks with Alice,} A more evolved back-and-forth sharing of packages that accounts for any previous multiversion replay decisions to be persisted, will require further changes both to the system and the algorithm. In such a scenario, if the caches persist, some intermediate results are available for free and the algorithm needs to accommodate for that accordingly. This scenario is part of our future work. 

Finally, our experiments show that \toolname significantly decreases the replay time for compute and data-intensive notebooks and allows a user to execute far higher number of versions in a given amount of time. The benefit arises particularly for notebooks where pre-processing or training steps are compute and data-intensive. In particular, if all computation is conducted in the last cell, then opportunities for optimization on intermediate results reduce drastically. In this case, one option is to further divide the last cell to look for opportunities of optimization. However, a better approach may be to look for opportunities within referenced function or libraries, if any.  







}

\vspace{-10pt}
\section{Conclusion}
\label{sec:conclusion}
In this work we have highlighted the need for improving the efficiency of multiversion replay. Our work shows that execution lineage can be used to establish cell equality and reuse shared program state to optimize replaying of multiversions. We show that optimizing is not trivial and,  
given a fixed cache size,  \probname\, is NP-hard and present two efficient heuristics for reducing the total computation time. We develop novel checkpoint-based caching support for replaying versions and show that \texttt{\TOOLNAME} is able to reduce the compute time of several machine learning and scientific computing notebooks using a cache size that is smaller than the checkpoint size of a notebook.\delete{, and in the same amount of time doubles the numbers of versions replayed by doubling the cache space.}

In the future, we wish to extend  
\texttt{\TOOLNAME} for queries and the standard database provenance model. This problem seems akin to how we previously extended provenance-based application virtualization~\cite{Pham:2013:PTU} to database virtualization~\cite{Pham:ICDE:LDV}.  We also wish to explore how \texttt{\TOOLNAME} can incorporate \revision{program restructuring}, which happens during interactive notebook development leveraging recent provenance models developed in this area~\cite{macke2020fine,koop2017dataflow,brachmann2020your} and developing corresponding online algorithms. 
\vspace{-5pt}
\begin{acks}
 This work is supported by National Science Foundation under grants CNS-1846418, NSF ICER-1639759, ICER-1661918 and a Department of Energy Fellowship. 
\end{acks}


\bibliographystyle{ACM-Reference-Format}
\bibliography{./main,./bagchi,./gehani,./paper.bib}

\pagebreak

\end{document}